\documentclass[fleqn,usenatbib]{mnras}

\usepackage[T1]{fontenc}

\DeclareRobustCommand{\VAN}[3]{#2}
\let\VANthebibliography\thebibliography
\def\thebibliography{\DeclareRobustCommand{\VAN}[3]{##3}\VANthebibliography}

\usepackage{graphicx}	
\usepackage{amsmath}	
\usepackage{amssymb}	

\usepackage[dvipsnames]{xcolor}
\usepackage{subcaption}

\usepackage{multirow}

\newcommand{\usmg}{USMg\textsc{ii} }

\newcommand{\OII}{[\mbox{O\,{\sc ii}}]}
\newcommand{\OIII}{[\mbox{O\,{\sc iii}}]}

\newcommand{\MgII}{\mbox{Mg\,{\sc ii}}}
\newcommand{\MgI}{\mbox{Mg\,{\sc i}}}

\newcommand{\FeII}{\mbox{Fe\,{\sc ii}} }
\newcommand{\MnII}{\mbox{Mn\,{\sc ii}} }
\newcommand{\CaII}{\mbox{Ca\,{\sc ii}} }

\usepackage{tabularx}

\newcommand\clearrow{\global\let\rowmac\relax}
\clearrow

\title[Nebular emissions associated with \MgII\ absorptions]{Strong nebular emissions associated with \MgII\ absorptions detected in the SDSS spectra of background quasars }

\author[Guha et al.]{
Labanya Kumar Guha,$^{1}$\thanks{E-mail: labanya@iucaa.in (LG)}
Raghunathan Srianand $^{1}$
\\
$^{1}$Inter-University Centre for Astronomy \& Astrophysics, Pune, 411007, India
}

\date{Accepted XXX. Received YYY; in original form ZZZ}

\pubyear{2021}

\begin{document}
\label{firstpage}
\pagerange{\pageref{firstpage}--\pageref{lastpage}}
\maketitle

\begin{abstract}
We present long-slit spectroscopic observations of { 40} Galaxy On Top Of Quasars (GOTOQs)  at $ {0.37 \leqslant z \leqslant 1.01}$ using the South African Large Telescope. Using this and available photometric data, we measure the impact parameters of the foreground galaxies to be in the range of 3$-$16 kpc with a median value of  { 8.6} kpc. This is the largest sample of galaxies producing \MgII\ absorption at such low impact parameters. These quasar-galaxy pairs are ideal for probing the disk-halo interface. At such impact parameters, we do not find any anti-correlation between rest equivalent width (REW) of Ca~{\sc ii}, Mn~{\sc ii}, Fe~{\sc ii}, Mg~{\sc ii} and Mg~{\sc i} absorption and impact parameter. These sight lines are typically redder than those of strong Mg~{\sc ii} absorbers, with { the color excess, E(B$-$V)} for our sample ranging from { $-$0.191} to { 0.422}, with a median value of { 0.058}.  In the E(B$-$V) vs. W$_{3935}$ plane, GOTOQs occupy the same region as Ca~{\sc ii} absorbers. For a given E(B$-$V), we find larger  W$_{3935}$ than what has been found in the Milky Way, probably due to a smaller dust-to-gas ratio in GOTOQs. Galaxy parameters could be measured for { twelve} cases, and their properties seem to follow the trends found for strong Mg~{\sc ii} absorbers.  Measuring the host galaxy properties for the full sample using HST photometry or AO-assisted ground-based imaging is important to gain insights into the relationship between the stellar mass of galaxies and the metal line REW  distributions at low impact parameters.
\end{abstract}

\begin{keywords}
galaxies: evolution – galaxies: groups: general – galaxies: haloes – galaxies: high-redshift – quasars: absorption
lines.
\end{keywords}



\section{Introduction}

Absorption lines imprinted in the spectra of bright background quasars are very useful in probing the physical conditions and the kinematics of the gas in the galactic disk and/or the circumgalactic medium (CGM) of foreground galaxies.  In particular, absorption produced by the \MgII$\rm{\lambda\lambda\, 2796,2803}$ doublet has been used to identify intervening galaxies \citep{bergeron1991, steidel1995}. Systematic studies over the past couple of decades and recent advances in the IFU-based spectroscopy \citep[see for example,][]{bouche2012, Nielsen_2013, Schroetter2019, Dutta2020, Huang2021, Lundgren2012, Lundgren2021}, have allowed us to establish the relationship between \MgII\ absorbers and host galaxies (or group of galaxies) over a wide range of impact parameters (D). However, these studies mostly probe large impact parameters, and we have very little statistics on the properties of \MgII\ absorbers at, say, $D\lesssim10$ kpc \citep[see][however]{Kacprzak_2013}. For example, out of about 560 unique quasar-galaxy pairs studied in various big surveys of \MgII\ absorption like MAGIICAT \citep{Nielsen_2013}, M3 halo project \citep{Huang2021} and MAGG \citep{Dutta2020}, only 5 have an impact parameter of less than ten kpc. \citet{kacprzak2012} have reported seven \MgII\ systems with $D < 10$ kpc, but most of these quasar-galaxy paris are restricted to very low redshifts (i.e., $z \sim 0.1$).
 
A quasar line of sight passing very close to the center of an intervening galaxy (say with $\rm{D \lesssim 10\, kpc}$) is most likely to probe the disk or disk-halo (or CGM) interface of the galaxy. Over the past few years, there have been significant explorations of the disk-halo interface of our own Galaxy \citep{Bish2019, Hobbs2020} as well as some of the low-redshift galaxies \citep{Boettcher2019}. These studies confirm the presence of inflows and outflows along the disk-halo interface consistent with the `galactic fountain' models \citep{Shapiro1991, Rubin2022}. Probing gas flows at such scales is important for understanding the sustained star formation in galaxies.

The best strategy for identifying  high-$z$ quasar-galaxy pairs with low impact parameters is to look for the nebular emission lines from the intervening galaxy in the spectrum of the background quasar. If a star-forming foreground galaxy is present within an angular separation of $1^{\prime \prime}$ (for example, in the case of SDSS DR-12) or $1.5^{\prime\prime}$ (for example in the case of SDSS DR-7) to the background quasar, the fiber used to observe the background quasar will pick the nebular line emissions from the foreground galaxy. This is now proven to be a very efficient technique for identifying host galaxies of \MgII\ absorption systems situated at very small angular separations \citep{Noterdaeme_2010, york2012, Straka_2015, Joshi2018, Kulkaerni2022}. Such systems are referred to as Galaxies On Top of Quasars \citep[or GOTOQs in short - ][]{york2012}. Starting from the \MgII$\rm{\lambda\lambda\, 2796,2803}$ absorption doublet present on the background quasar spectra, \citet{Joshi2017} have identified 198 GOTOQs over the redshift range $0.35 \leqslant z \leqslant 1.1$. However, note that measuring the impact parameters of the foreground galaxies directly from the fiber-fed spectroscopy is not possible.

In \citet{Guha2023}, we have investigated the nature of these GOTOQs using multiband images available from the Dark Energy Spectroscopic Instrument (DESI) Legecy Survey\citep{Dey2015, Dey2019}. There are 74 GOTOQs in the sample of \citet{Joshi2017} for which photometric extension around the quasar image was detected in deep images, after subtracting the quasar contribution \citet{Guha2023} were able to measure the impact parameters of the galaxies to be in the range $5.9\leqslant D[kpc]\leqslant 16.9$. This has substantially increased the number of \MgII\ absorbing host galaxies probed at such low impact parameters and allowed us to measure $W_{2796}(D=0)$ of $3.44\pm0.20$ \AA\ and an exponential scale length of 21.6$^{+2.41}_{-1.97}$ kpc for the $\rm W_{2796} - D$ relationship. While the photometric redshift estimated for the residual light (after subtracting the quasar contribution) is consistent (within $\rm 2\sigma$) with the redshift of the \MgII\ absorbers, independent confirmation using the spatially resolved galaxy spectra is unavailable.

One of the primary objectives of this study is to verify that the photometric extensions observed in GOTOQs correspond to the host galaxies of \MgII\ absorption. Another aim is to identify the locations of galaxies in cases where there are no clear extensions visible in the available photometric images. We use long-slit spectroscopy using the Southern African Large Telescope \citep[SALT,][]{buckley_2005} with slit orientated along different position angles for this purpose.

This paper is organized as follows. Section \ref{sec:sample} describes the sample constructed to study the host galaxies of strong \MgII\ absorption systems at small impact parameters, and section \ref{sec:observations} describes settings used for our SALT observations and data reduction. Section \ref{sec:results} describes our main results, while section \ref{sec:discussions} summarizes the discussions and conclusions of this paper. Throughout this paper, we use a flat $\rm{\Lambda CDM}$ cosmology with Hubble constant, $H_0$ = 70 $\rm{km s^{-1} Mpc^{-1}}$ and $\rm \Omega_m$ = 0.3.

\section{The GOTOQS sub-sample}
\label{sec:sample}

\begin{table*}
    \centering
    \begin{tabular}{>{\rowmac}l>{\rowmac}l>{\rowmac}c>{\rowmac}c>{\rowmac}c>{\rowmac}c>{\rowmac}c>{\rowmac}c>{\rowmac}c>{\rowmac}l>{\rowmac}l<{\clearrow}}
\hline
(1) & (2) & (3) & (4) & (5) & (6) & (7) & (8) & (9) & (10) & (11) \\
No. & Quasar & Plate & MJD & Fiber & $z_{qso}$ & $z_{abs}$ & $z_{ems}$ & $\rm{L_{\text{[}O\, \textsc{ii}\text{]}}}$ &  $\rm{W_{2796}\, \text{(\AA)}}$  & E(B$-$V$)_{\rm SMC}$\\
\hline
\hline
1   & BOSS-J000123.61$-$032317.9   & 4354 & 55810 & 0282 & 3.1210 & 0.8918 & 0.8922 & 33.67  $\pm$ 4.75 & 2.12 $\pm$ 0.28 & $+$0.062$\pm$0.004 \\
2   & BOSS-J005436.35+005055.6     & 4224 & 55481 & 0746 & 2.2363 & 0.4810 & 0.4812 & 14.36  $\pm$ 0.74 & 1.87 $\pm$ 0.37 & $+$0.090$\pm$0.003 \\
    & SDSS-J005436.35+005055.6     & 0692 & 52201 & 0539 & 2.2386 & 0.4806 & 0.4813 & 15.55  $\pm$ 2.56 & 1.91 $\pm$ 0.75 & $+$0.084$\pm$0.002 \\
3   & SDSS-J005408.46$-$094638.2   & 0657 & 52177 & 0515 & 2.1232 & 0.4778 & 0.4778 & 15.81  $\pm$ 2.30 & 2.41 $\pm$ 0.15 & $+$0.076$\pm$0.001 \\  
4   & BOSS-J012916.14$-$003211.4   & 4229 & 55501 & 0182 & 1.8370 & 0.8991 & 0.8997 & 31.44  $\pm$ 3.30 & 3.50 $\pm$ 0.26 & $+$0.048$\pm$0.002 \\
5   & BOSS-J013132.92$-$040137.5   & 7048 & 56575 & 0710 & 2.2700 & 0.5017 & 0.5017 & 17.78  $\pm$ 2.56 & 1.70 $\pm$ 0.12 & $+$0.085$\pm$0.002 \\
6   & SDSS-J020317.22$-$010122.8   & 0701 & 52179 & 0087 & 1.7210 & 0.7253 & 0.7258 & 26.10  $\pm$ 7.53 & 2.10 $\pm$ 0.51 & $+$0.050$\pm$0.005 \\
7   & BOSS-J023529.82+005916.3     & 3744 & 55209 & 0952 & 2.7160 & 0.9250 & 0.9245 & 34.07  $\pm$ 5.42 & 0.93 $\pm$ 0.43 & $+$0.051$\pm$0.006 \\
8   & BOSS-J023608.18+005604.0     & 6781 & 56274 & 0791 & 0.9680 & 0.8212 & 0.8216 & 19.70  $\pm$ 2.55 & 2.02 $\pm$ 0.12 & $+$0.011$\pm$0.002\\
    & SDSS-J023608.18+005604.0     & 0408 & 51821 & 0321 & 0.9671 & 0.8212 & 0.8216 & 10.67  $\pm$ 3.56 & 1.77 $\pm$ 0.27 & $+$0.037$\pm$0.003\\
9   & SDSS-J023618.98$-$000529.1   & 0408 & 51821 & 0237 & 0.9806 & 0.6448 & 0.6460 & 17.16  $\pm$ 4.86 & 1.26 $\pm$ 0.40 & $+$0.079$\pm$0.004\\ 
10  & SDSS-J030730.61$-$074555.6   & 0459 & 51924 & 0309 & 0.7543 & 0.6986 & 0.6994 & 28.81  $\pm$ 4.00 & 2.34 $\pm$ 0.27 & $+$0.100$\pm$0.002\\
11  & BOSS-J080751.03+094554.7     & 4510 & 55559 & 0081 & 2.5900 & 0.7470 & 0.7470 & 24.17  $\pm$ 2.79 & 2.12 $\pm$ 0.21 & $-$0.035$\pm$0.003\\
12  & SDSS-J080808.67+064108.7     & 1756 & 53080 & 0184 & 2.1133 & 0.4326 & 0.4326 & 8.96   $\pm$ 2.31 & 1.96 $\pm$ 0.59 & $+$0.007$\pm$0.003\\  
13  & SDSS-J084915.67+065013.0     & 1189 & 52668 & 0322 & 2.1970 & 0.8432 & 0.8437 & 140.24 $\pm$ 4.25 & 2.60 $\pm$ 0.14 & $+$0.056$\pm$0.004\\
    & BOSS-J084915.67+065013.0     & 5289 & 55893 & 0128 & 2.1970 & 0.8432 & 0.8438 & 125.55 $\pm$ 6.90 & 3.00 $\pm$ 0.27 & $+$0.042$\pm$0.004\\
14  & {SDSS-J085051.97+083026.6}     & 1299 & 52972 & 0333 & 0.6635 & 0.3767 & 0.3767 & 5.56   $\pm$ 1.74 & 1.75 $\pm$ 0.15 &  $+$0.054$\pm$0.003\\             
15  & SDSS-J085830.53+055606.0     & 1191 & 52674 & 0283 & 1.8524 & 0.4318 & 0.4318 & 17.76  $\pm$ 1.62 & 2.90 $\pm$ 1.03 & $+$0.052$\pm$0.004 \\
16  & {SDSS-J090830.32+042436.3}     & 1192 & 52649 & 0019 & 1.7518 & 0.4362 & 0.4363 & 5.02   $\pm$ 2.01 & 1.69 $\pm$ 0.37 & $-$0.001$\pm$0.002\\
17  & {BOSS-J091709.35+033734.0}     & 3820 & 55542 & 0008 & 2.1060 & 0.6622 & 0.6620 & 18.82  $\pm$ 4.78 & 2.66 $\pm$ 0.80 & $+$0.030$\pm$0.003\\
18  & BOSS-J092915.02-012621.0     & 3781 & 55243 & 0978 & 2.2500 & 0.6990 & 0.7002 & 8.57   $\pm$ 1.85 & 1.47 $\pm$ 0.43 & $-$0.191$\pm$0.004\\
19  & {SDSS-J094222.51+001641.9}     & 0266 & 51630 & 0396 & 1.4065 & 0.5243 & 0.5247 & 20.86  $\pm$ 2.80 & 2.44 $\pm$ 0.23 & $+$0.100$\pm$0.002\\ 
20  & BOSS-J101446.63+083500.9     & 5334 & 55928 & 0860 & 2.5120 & 1.0017 & 1.0012 & 49.27  $\pm$ 8.38 & 4.75 $\pm$ 0.08 & $+$0.073$\pm$0.002\\
    & SDSS-J101446.64+083500.9     & 3155 & 54829 & 0179 & 2.5113 & 1.0017 & 1.0011 & 46.53  $\pm$ 8.43 & 4.39 $\pm$ 0.10 & $+$0.093$\pm$0.003\\
21 & BOSS-J102518.88$-$001755.6    & 3832 & 55289 & 0388 & 2.2850 & 0.8801 & 0.8803 & 39.53  $\pm$ 6.84 & 3.82 $\pm$ 0.29 & $+$0.107$\pm$0.002 \\
22 & {BOSS-J111755.71+075932.2}      & 5369 & 56272 & 0393 & 2.6350 & 0.7681 & 0.7687 & 24.01  $\pm$ 2.72 & 1.98 $\pm$ 0.61 &  $+$0.012$\pm$0.005\\ 
23 & SDSS-J120908.53+022734.3      & 0517 & 52024 & 0541 & 1.3370 & 0.6695 & 0.6695 & 59.08  $\pm$ 3.97 & 3.27 $\pm$ 0.22 & $+$0.033$\pm$0.003 \\
24 & {SDSS-J121242.26+083410.9}      & 1624 & 53386 & 0453 & 1.5368 & 0.5330 & 0.5322 & 12.56  $\pm$ 4.88 & 2.12 $\pm$ 0.20 & $+$0.018$\pm$0.002\\
25 & {BOSS-J121838.10+030900.1}      & 4750 & 55630 & 0774 & 2.2510 & 0.5222 & 0.5220 & 11.53  $\pm$ 1.85 & 2.80 $\pm$ 0.39 & $-$0.014$\pm$0.002\\   
26 & BOSS-J123514.36+030416.7      & 4754 & 55649 & 0592 & 2.1950 & 0.5313 & 0.5338 & 17.82  $\pm$ 2.37 & 1.36 $\pm$ 0.23 & $+$0.125$\pm$0.002 \\       
27 & {BOSS-J131454.20+033042.3}      & 4760 & 55656 & 0096 & 2.7000 & 0.7792 & 0.7801 & 18.72  $\pm$ 2.71 & 1.84 $\pm$ 0.21 & $+$0.422$\pm$0.003\\
28 & BOSS-J131918.88$-$030621.2    & 4049 & 55591 & 0418 & 2.3400 & 0.6772 & 0.6777 & 32.77  $\pm$ 6.30 & 2.06 $\pm$ 0.47 & $+$0.060$\pm$0.003\\
29 & SDSS-J135445.72+075637.6      & 1806 & 53559 & 0043 & 1.3235 & 0.7126 & 0.7119 & 26.39  $\pm$ 3.95 & 2.15 $\pm$ 0.31 & $+$0.095$\pm$0.003\\
30 & SDSS-J135634.16+060133.6      & 1805 & 53875 & 0013 & 1.3929 & 0.6292 & 0.6298 & 29.37  $\pm$ 10.55& 2.52 $\pm$ 0.19 & $-$0.007$\pm$0.002\\
31 & SDSS-J140446.28+062340.2      & 1824 & 53491 & 0231 & 1.1852 & 0.4515 & 0.4521 & 16.11  $\pm$ 1.85 & 2.28 $\pm$ 0.33 & $+$0.142$\pm$0.002\\
32 & BOSS-J142232.38$-$003043.9    & 4028 & 55621 & 0422 & 2.7060 & 1.0127 & 1.0127 & 40.50  $\pm$ 13.70& 1.59 $\pm$ 0.13 & $+$0.014$\pm$0.003\\
33 & SDSS-J142953.20+041551.6      & 0585 & 52027 & 0473 & 1.9410 & 0.7134 & 0.7126 & 22.40  $\pm$ 4.21 & 3.57 $\pm$ 0.33 & $+$0.116$\pm$0.003 \\
34 & SDSS-J161728.95+061604.3      & 1731 & 53884 & 0113 & 1.2501 & 0.7884 & 0.7878 & 90.74  $\pm$ 5.72 & 3.20 $\pm$ 0.18 & $+$0.097$\pm$0.002\\
35 & BOSS-J215059.09+022557.2      & 5146 & 55831 & 0878 & 2.3780 & 0.5682 & 0.5673 & 7.01   $\pm$ 1.45 & 2.37 $\pm$ 0.92 & $+$0.052$\pm$0.003\\
36 & BOSS-J223244.93+011058.1      & 4203 & 55447 & 0804 & 1.0090 & 0.4899 & 0.4901 & 4.20   $\pm$ 0.96 & 1.69 $\pm$ 0.37 & $+$0.185$\pm$0.002\\
37 & BOSS-J224405.09$-$021547.9    & 4363 & 55537 & 0778 & 2.5740 & 0.9243 & 0.9246 & 18.73  $\pm$ 2.92 & 3.77 $\pm$ 0.20 & $+$0.088$\pm$0.002 \\
38 & BOSS-J225028.85+031146.4      & 4293 & 55509 & 0453 & 2.5340 & 0.7198 & 0.7196 & 17.31  $\pm$ 2.97 & 2.67 $\pm$ 0.12 & $+$0.012$\pm$0.002\\
39 & SDSS-J234257.50+010121.3      & 1903 & 53357 & 0420 & 2.4520 & 0.6767 & 0.6764 & 11.24  $\pm$ 2.78 & 3.10 $\pm$ 0.38 & $+$0.057$\pm$0.004\\
   & BOSS-J234257.50+010121.3      & 4213 & 55449 & 0912 & 2.4520 & 0.6768 & 0.6763 & 8.45   $\pm$ 3.23 & 2.35 $\pm$ 0.27 & $+$0.074$\pm$0.003 \\
40 & BOSS-J235639.31$-$040614.4    & 7034 & 56564 & 0506 & 2.8800 & 0.7709 & 0.7705 & 55.89  $\pm$ 8.51 & 3.63 $\pm$ 0.34 & $+$0.062$\pm$0.005\\

\hline

    \end{tabular}
    \caption{Details of the \MgII\ absorption systems constituting the GOTOQs sub-sample for this work. The second column provides the name of the GOTOQ. The third, fourth, and fifth columns provide the plate, MJD, and fiber IDs of the SDSS observation of the GOTOQ. Columns (6), (7), and (8) provide the redshift of the background quasar, intervening \MgII\ absorption, and the foreground galaxy, respectively. The \OII\ line luminosity detected within the SDSS fiber is given in column (9).The \OII\ luminosities are given in the units of $\rm 10^{40}\, ergs\, s^{-1}$. The rest equivalent width of the \MgII\ $\lambda$ 2796 line is given in column (10). Column (11) provides the color excess due to the line of sight reddening of the quasar due to the foreground galaxy.}
    \label{tab:sample}
\end{table*}

\begin{figure*}

\begin{minipage}{0.36\linewidth}
\centering
\includegraphics[width=0.98\textwidth]{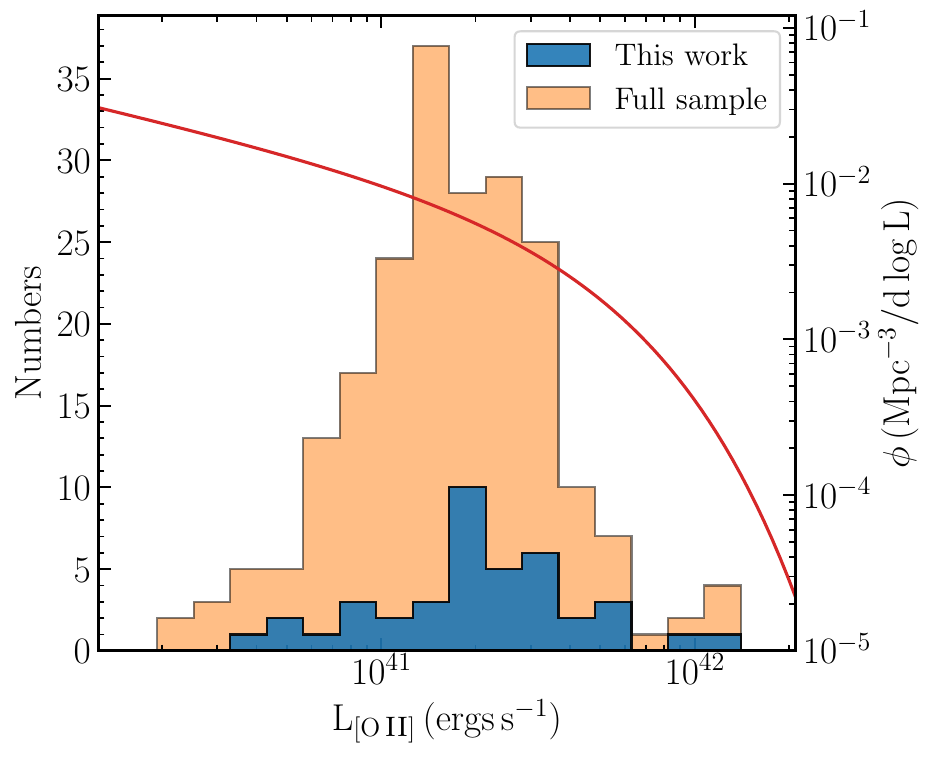}
\end{minipage}
\begin{minipage}{0.315\linewidth}
\centering
\includegraphics[width=0.95\textwidth]{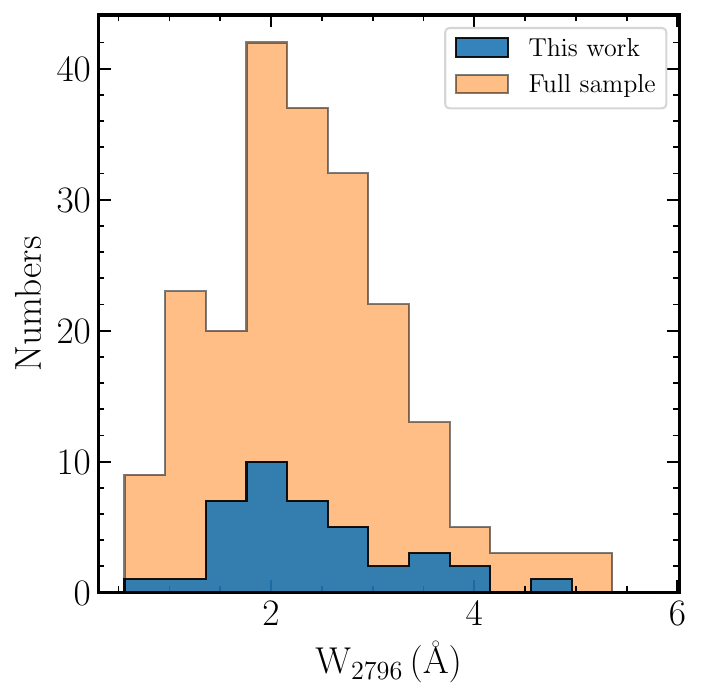}
\end{minipage}
\begin{minipage}{0.315\linewidth}
\centering
\includegraphics[width=0.95\textwidth]{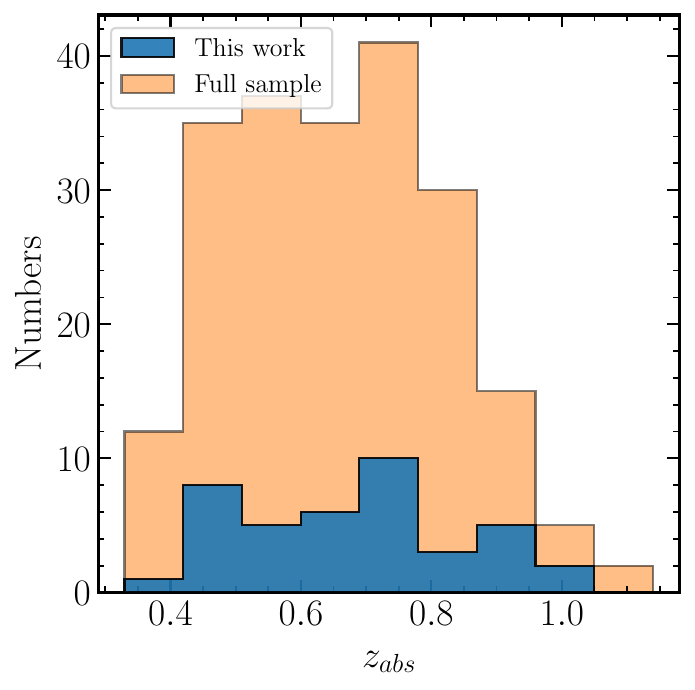}
\end{minipage}

\caption{Comparison between the GOTOQs sub-sample considered for this work and the full GOTOQs sample constructed by \citet{Joshi2017} based on various absorption and emission line properties.  The left panel shows the histograms of \OII\ line luminosity detected on top of the quasar spectrum associated with the \MgII\ absorption for our GOTOQs sub-sample (blue) and the entire sample (orange).  The solid red line corresponds to the Schechter fit of the \OII\ line luminosity function \citep{Comparat2016} at $z = 0.69$, the median redshift of our GOTOQs sub-sample. Similarly, the middle panel shows the histogram of $\rm{W_{2796}}$ while the right panel shows the histogram of the \MgII\ absorption redshift.}
\label{fig:hist_comparison}
\end{figure*}

In this work, we mainly focus on a sub-sample of GOTOQs from the original sample of \citet{Joshi2017}. In brief, \citet{Joshi2017} inspected all the quasar spectra with intervening \MgII\ absorption listed in the SDSS \FeII/\MgII\ metal absorber catalog \citep{Zhu2013} and searched for associated nebular emissions on top of the quasar spectra, utilizing either the \OII\ line with a minimum detection significance of $\rm{4\sigma}$ or each of the \OIII\ $\lambda\lambda$ 4960, 5008 lines with $\rm{3\sigma}$ significances. This resulted in 198 GOTOQs in the redshift interval 0.35 $\leqslant z_{abs} \leqslant$ 1.1. As our primary focus of this study is to precisely calculate the impact parameter of the foreground \MgII\ absorption host galaxies using the long-slit spectra obtained at different position angles (PAs), while constructing the GOTOQs sub-sample for this work, we only considered systems that are accessible to the SALT, i.e., the declination of the systems being less than $\rm{10}^{\circ}$.

We introduced a cut-off on the \OII\ luminosity to limit the number of sources. We only consider those systems for which the luminosity \OII\ emission line detected in SDSS quasar spectra are at least  ${\rm{{L_{\lbrack O\,\textsc{ii}\rbrack}^{thres}}}}$ = $\rm{{4.2\times 10^{40}\, ergs\, s^{-1}}}$. Given the SFR calibrations \citep{Kennicutt1998SFR}, this threshold in \OII\ line luminosity corresponds to a minimum SFR (uncorrected for dust extinction and fiber loss) of $\sim\rm{{0.6\, M_{\odot}\, yr^{-1}}}$. However, note that the actual SFR can be much higher since we did not correct the fiber effects and the extinctions due to interstellar dust. The two criteria discussed above resulted in { 53} GOTOQs, out of which we completed long-slit observations for { 40} systems  using SALT. The details of the  GOTOQs observed with SALT are presented in Table~\ref{tab:sample}.  The \MgII\ rest equivalent width, $\rm W_{2796}$ of the observed sample varies between { 0.93\AA\ to 4.75\AA\ with a median value of { 2.21\AA}}. 

To compare this GOTOQs sub-sample against the original sample, in Figure \ref{fig:hist_comparison}, we plot the histograms of both these systems based on various absorption and emission line properties. The left panel shows the histograms of \OII\ line luminosity for our GOTOQs sub-sample (blue) and the entire sample (orange). By construction, this sub-sample should correspond to the more luminous \OII\ systems, as evident from this plot. For the original sample \OII\ luminosity ranges $\rm{{1.94\, \leqslant L_{[O\, \textsc{ii}]}\, [10^{40}\, ergs\, s^{-1}]\, \leqslant\, 140.24}}$ with a median value of $\rm{{16.7 \times\, 10^{40}\, ergs\, s^{-1}}}$ while for the sub-sample the \OII\ luminosity ranges  $\rm{{4.2\, \leqslant L_{[O\, \textsc{ii}]}\, [10^{40}\, ergs\, s^{-1}]\leqslant\, 140.24}}$ with a median value of $\rm{{19.3 \times\, 10^{40}\, ergs\, s^{-1}}}$. 

The middle panel shows the histograms of $\rm{W_{2796}}$ for our GOTOQs sub-sample (blue) and the entire sample (orange). Compared to the $\rm{W_{2796}}$ range of $\rm{{0.62 \leqslant W_{2796}[\text{\AA}] \leqslant 5.33}}$ with a median value of 2.38\AA\ for the full sample, for our sample $\rm{{W_{2796}}}$ ranges $\rm{{0.93 \leqslant W_{2796}[\text{\AA}] \leqslant 4.75}}$ with a median value of { 2.21\AA}. The plot shows that the distributions of $\rm{{W_{2796}}}$ for the two samples are similar. The KS test yields that these two $\rm{{W_{2796}}}$ distributions are drawn from the same underlying population ({ with a  $p$ value of $\sim$ 0.85}).

Finally, the right panel shows the histograms of the \MgII\ absorption redshift for our GOTOQs sub-sample (blue) and the complete sample (orange). The absorption redshift for our sample and the complete sample varies over the range { 0.37 $\leqslant\, z_{abs}\, \leqslant$ 1.01 with a median value of 0.69} and 0.36 $\leqslant\, z_{abs}\, \leqslant$ 1.1 with a median value of 0.66 respectively. KS test yields that the redshift distributions of these two samples are likely drawn from the same distributions ({ with a $p$ value of $\sim$ 0.90}). {\it In summary, the GOTOQ sub-sample used in this work probes the GOTOQs systems over the complete $\rm{W_{2796}}$ and redshift ranges with slightly higher $\rm{L_{[O~\textsc{ii}]}}$ compared to the primary GOTOQs sample.}

\section{Observations and data reductions}
\label{sec:observations}

\begin{table*}
    \centering
    \begin{tabular}{>{\rowmac}c>{\rowmac}c>{\rowmac}c>{\rowmac}c>{\rowmac}c>{\rowmac}c>{\rowmac}c<{\clearrow}}
    \hline
    No. & Quasar   &  Date & Exposure (s)  &  PA (deg.) & Grating Angel (deg.)    &  Wavelength Range (\AA)  \\
    (1) & \multicolumn{1}{c}{(2)} & (3) & (4) & (5) & (6) & (7) \\
    \hline
    \hline
    1    &  J000123.61$-$032317.9 & 2017-06-26 & 2480 &  90  & 21.5    & 6560 -- 9510 \\
         &                        & 2017-07-01 & 2480 &  90  & 21.5    & 6560 -- 9510 \\
         &                        & 2023-07-17 & 2480 &  150 & 21.125  & 6420 -- 9370 \\
    2    &  J005408.46$-$094638.2 & 2018-08-12 & 2500 &  45  & 17.375  & 5050 -- 8050 \\
         &                        & 2018-10-19 & 2600 &  315 & 17.375  & 5050 -- 8050 \\
    3    &  J005436.35+005055.6   & 2023-07-17 & 2500 &  75  & 17.0    & 4900 -- 7900\\ 
    4    &  J012916.14$-$003211.4 & 2017-07-27 & 2480 &  315 & 21.5    & 6560 -- 9510 \\
         &                        & 2017-08-28 & 2480 &  315 & 21.5    & 6560 -- 9510 \\
    5    &  J013132.92$-$040137.5 & 2023-08-04 & 2500 &  15  & 17.375  & 5050 -- 8050 \\
    6    &  J020317.22$-$010122.8 & 2017-08-03 & 2500 &  135 & 19.625  & 5860 -- 8860 \\
         &                        & 2017-09-13 & 2480 &  45  & 19.625  & 5860 -- 8860 \\
    7    &  J023529.82+005916.3   & 2017-07-29 & 2520 &  60  & 21.5    & 6560 -- 9510 \\
         &                        & 2017-10-17 & 2400 &  330 & 21.5    & 6560 -- 9510 \\
    8    &  J023608.18+005604.0   & 2023-09-19 & 2480 &  90  & 20.75   & 6280 -- 9250 \\
         &                        & 2023-09-12 & 2480 &  0   & 20.75   & 6280 -- 9250 \\
    9    &  J023618.98$-$000529.1 & 2023-10-16 & 2500 &  100 & 18.875  & 5590 -- 8590 \\
    10   &  J030730.61$-$074555.6 & 2022-10-27 & 2400 &  0   & 19.625  & 5860 -- 8860 \\
         &                        & 2022-10-01 & 2400 &  90  & 19.625  & 5860 -- 8860 \\
    11   &  J080751.03+094554.7   & 2023-12-19 & 2200 &  80  & 20.0    & 6005 -- 8985 \\
         &                        & 2023-12-22 & 2200 &  170 & 20.0    & 6005 -- 8985 \\
    12   &  J080808.67+064108.7   & 2023-12-16 & 2300 &  90  & 16.625  & 4780 -- 7780 \\
    13   &  J084915.67+065013.0   & 2022-02-23 & 2300 &  63  & 21.125  & 6420 -- 9370 \\
         &                        & 2022-02-26 & 2300 &  162 & 21.125  & 6420 -- 9370 \\
    14   &  {J085051.97+083026.6}   & 2024-04-03 & 2240 &  45  & 16.25   & 4620 -- 7650 \\
    15   &  J085830.53+055606.0   & 2022-12-18 & 2560 &  112 & 16.625  & 4780 -- 7780 \\
         &                        & 2023-01-23 & 2400 &  22  & 16.625  & 4780 -- 7780 \\
    16   &  {J090830.32+042436.3}   & 2024-04-12 & 2240 &  135 & 16.625  & 4780 -- 7780 \\
    17   &  {J091709.35+033734.0}   & 2024-04-04 & 2500 &  90  & 18.875  & 5590 -- 8590 \\
    18   &  J092915.02$-$012621.0 & 2023-12-13 & 2500 &  90  & 19.25   & 5725 -- 8725 \\
    19   &  {J094222.51+001641.9}   & 2024-03-06 & 2500 &  90  & 17.75   & 5170 -- 8190 \\ 
    20   &  J101446.63+083500.9   & 2022-02-06 & 2240 &  177 & 19.625  & 5860 -- 8860 \\
    21   &  J102518.88$-$001755.6 & 2022-12-30 & 2560 &  0   & 21.5    & 6560 -- 9510 \\
         &                        & 2022-04-20 & 2540 &  77  & 21.5    & 6560 -- 9510 \\
    22   &  {J111755.71+075932.2}   & 2024-03-05 & 2240 &  315 & 20.0    & 6005 -- 8985 \\
    23   &  J120908.53+022734.3   & 2018-02-16 & 2400 &  0   & 19.25   & 5725 -- 8725 \\
         &                        & 2018-02-22 & 2400 &  90  & 19.25   & 5725 -- 8725 \\
    24   &  {J121242.26+083410.9}   & 2024-04-13 & 2200 &  0   & 17.375  & 5050 -- 8050 \\
    25   &  {J121838.10+030900.1}   & 2024-02-12 & 2500 &  315 & 17.75   & 5165 -- 8195 \\
    26   &  J123514.36+030416.7   & 2023-04-25 & 2560 &  0   & 17.75   & 5165 -- 8195 \\
    27   &  {J131454.20+033042.3}   & 2024-03-09 & 2500 &  90  & 20.375  & 6150 -- 9120 \\
         &                        & 2024-03-10 & 2100 &  0   & 20.375  & 6150 -- 9120 \\
    28   &  J131918.88$-$030621.2 & 2017-06-16 & 2500 &  45  & 19.25   & 5725 -- 8725 \\
         &                        & 2017-07-16 & 2335 &  315 & 19.25   & 5725 -- 8725 \\
    29   &  J135445.72+075637.6   & 2017-05-18 & 2200 &  315 & 19.625  & 5860 -- 8860 \\
         & 	                     & 2017-05-18 &	2200 &	45  & 19.625  & 5860 -- 8860 \\
    30   &  J135634.16+060133.6   & 2017-04-28 & 2200 &	 277 & 18.875  & 5590 -- 8590 \\
         &                        & 2017-05-20 & 2200 &  20  & 18.875  & 5590 -- 8590 \\
    31   &  J140446.28+062340.2   & 2023-07-17 & 2300 &  0   & 16.625  & 4780 -- 7780 \\
    32   &  J142232.38$-$003043.9 &	2017-05-29 & 2600 &  18  & 18.5    & 5450 -- 8450 \\
         &                        & 2017-07-20 &	2500 &	336 & 17.75   & 5170 -- 8190 \\
    33   &  J142953.20+041551.6   & 2022-05-02 & 2300 &  76  & 19.625  & 5870 -- 8840 \\
         &                        & 2022-05-03 & 2300 &  166 & 19.625  & 5870 -- 8840 \\
    34   &  J161728.95+061604.3   &	2017-04-29 & 2300 &	 45  & 20.75   & 6280 -- 9250 \\
         &                        & 2017-05-19 &	2137 &  330 & 20.75   & 6280 -- 9250 \\
    35   &  J215059.09+022557.2   & 2023-08-12 & 2500 &  175 & 18.125  & 5300 -- 8300 \\
    36   &  J223244.93+011058.1   & 2023-10-15 & 2500 &  130 & 17.0    & 4900 -- 7900 \\
    37   &  J224405.09$-$021547.9 & 2023-06-23 & 2500 &  165 & 21.5    & 6560 -- 9510 \\
    38   &  J225028.85+031146.4   & 2023-07-17 & 2480 &  5   & 19.625  & 5870 -- 8840 \\
    39   &  J234257.50+010121.3   & 2023-07-22 & 2300 &  135 & 18.875  & 5590 -- 8590 \\
    40   &  J235639.31$-$040614.4 & 2017-06-06 & 2500 &  45  & 20.375  & 6150 -- 9120 \\
         &                        & 2017-08-19 & 2350 &  315 & 20.375  & 6150 -- 9120 \\
         &                        & 2020-09-08 & 2560 &  90  & 20.0    & 6005 -- 8985 \\

    \hline
    \end{tabular}
    \caption{Observation log of the observed sub-sample. Column (2) represents the name of the GOTOQs. Columns (3), (4), and (5) correspond to the date of the observations, on-source exposure time, and the position angle used to observe the GOTOQs, respectively. Columns (6) and (7) correspond to the grating angle used and the associated wavelength ranges suitably chosen for detecting the nebular lines. }
    \label{tab:observations}
    \end{table*}

\begin{figure*}
  \begin{subfigure}{0.19\linewidth}
    \centering\includegraphics[width=\textwidth]{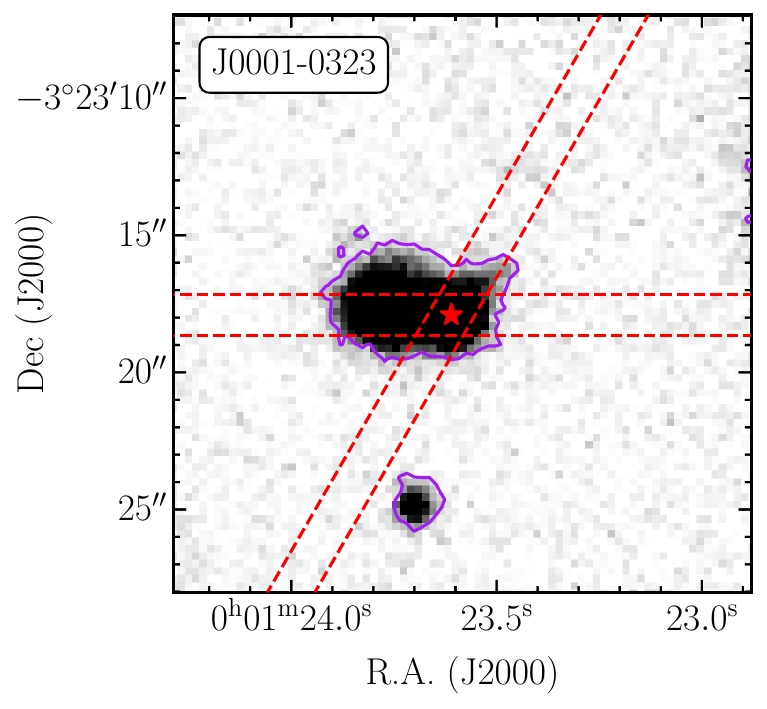}
  \end{subfigure}
  \begin{subfigure}{0.19\textwidth}
    \centering\includegraphics[width=\textwidth]{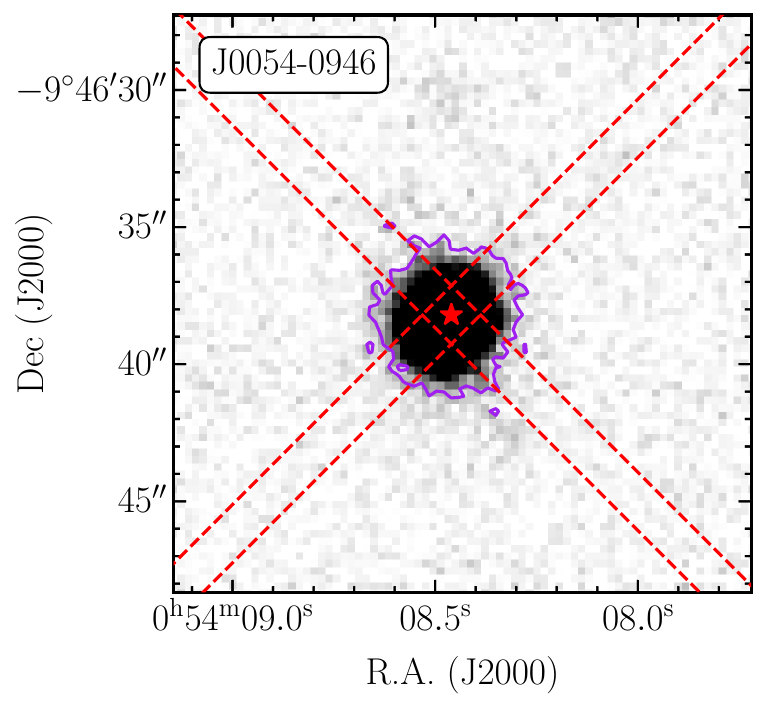}
  \end{subfigure}
  \begin{subfigure}{0.19\textwidth}
    \centering\includegraphics[width=\textwidth]{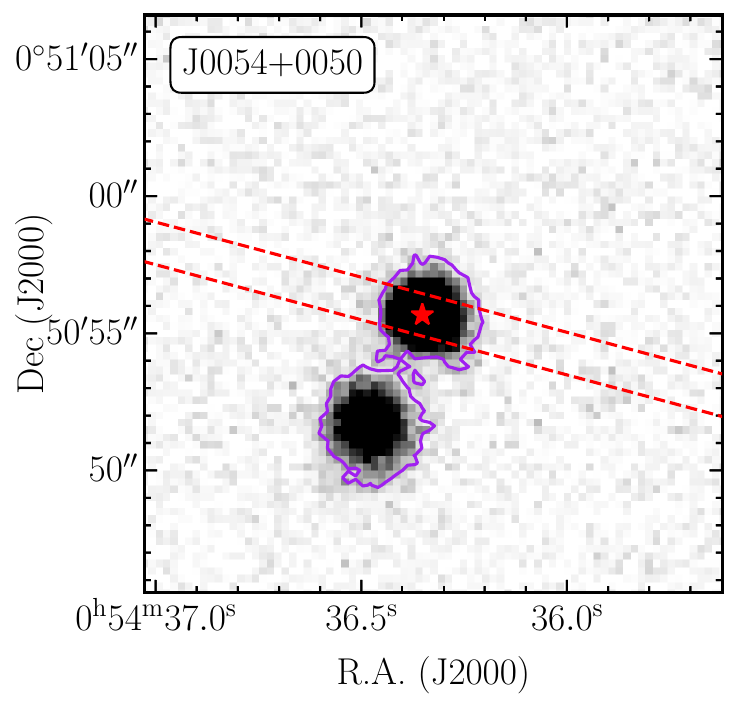}
  \end{subfigure}
  \begin{subfigure}{0.19\textwidth}
    \centering\includegraphics[width=\textwidth]{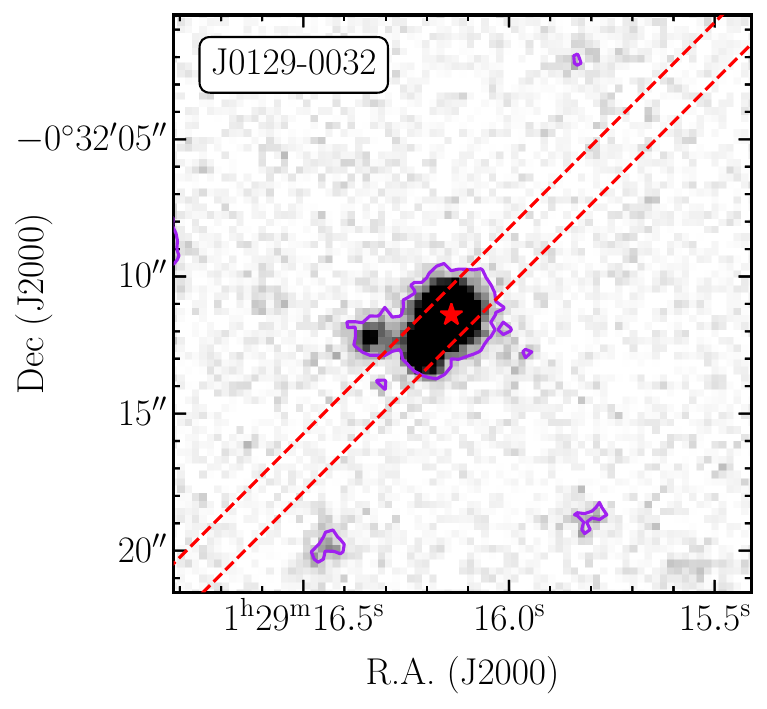}
  \end{subfigure}
  \begin{subfigure}{0.19\textwidth}
    \centering\includegraphics[width=\textwidth]{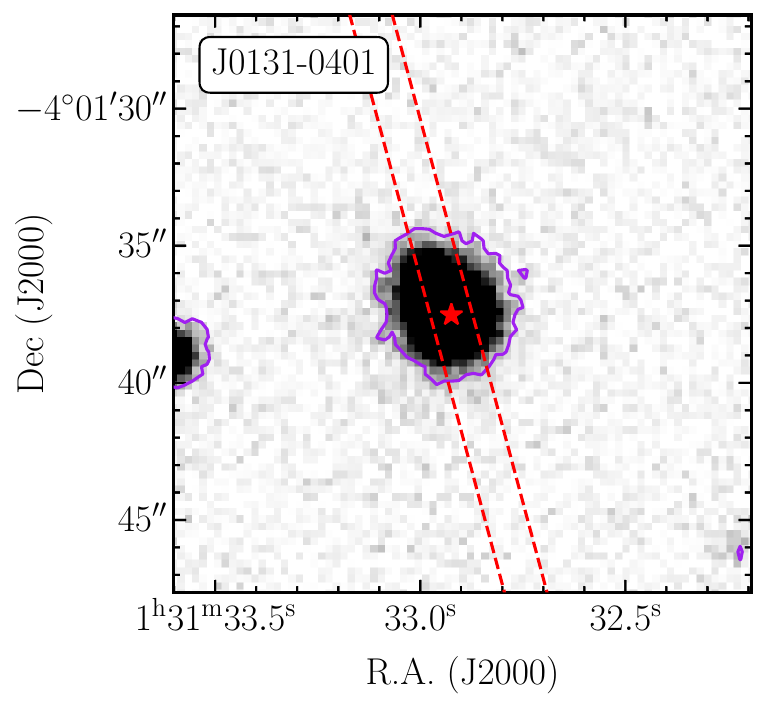}
  \end{subfigure}
  
  \begin{subfigure}{0.19\textwidth}
    \centering\includegraphics[width=\textwidth]{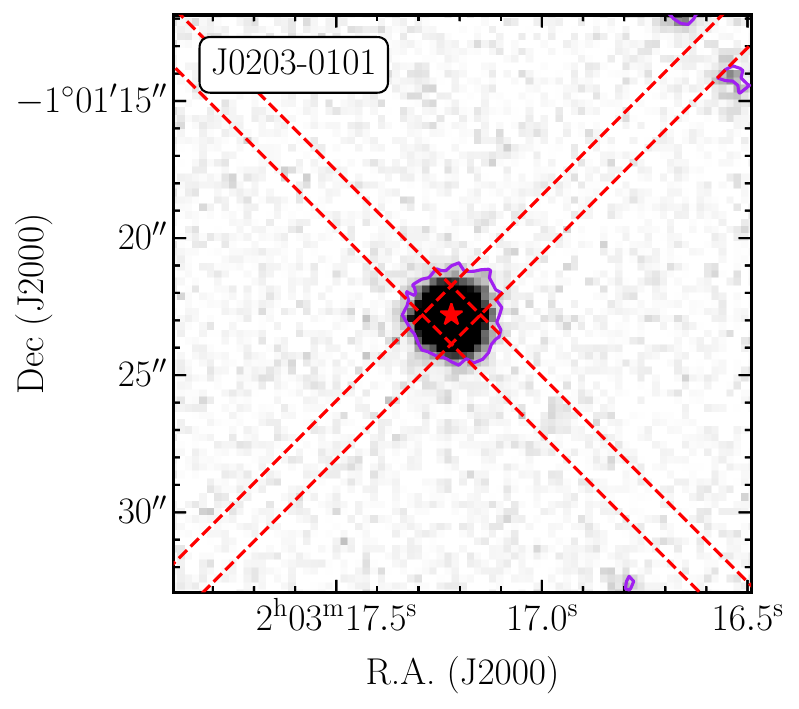}
  \end{subfigure}
  \begin{subfigure}{0.19\textwidth}
    \centering\includegraphics[width=\textwidth]{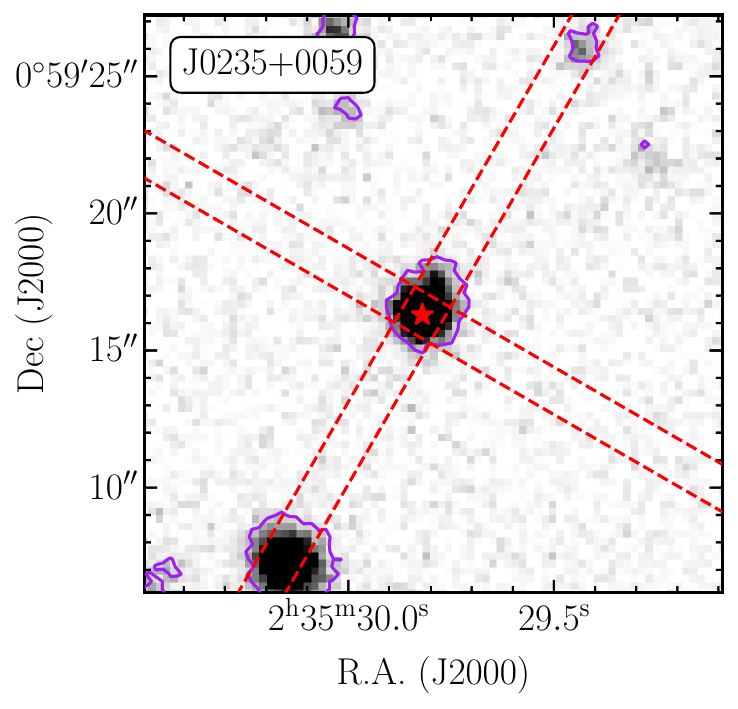}
  \end{subfigure}
  \begin{subfigure}{0.19\textwidth}
    \centering\includegraphics[width=\textwidth]{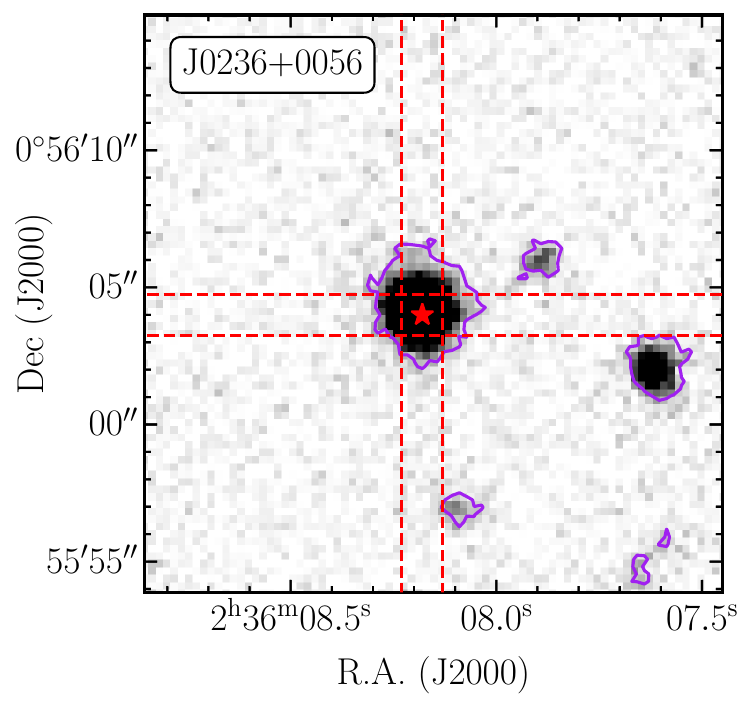}
  \end{subfigure}
  \begin{subfigure}{0.19\textwidth}
    \centering\includegraphics[width=\textwidth]{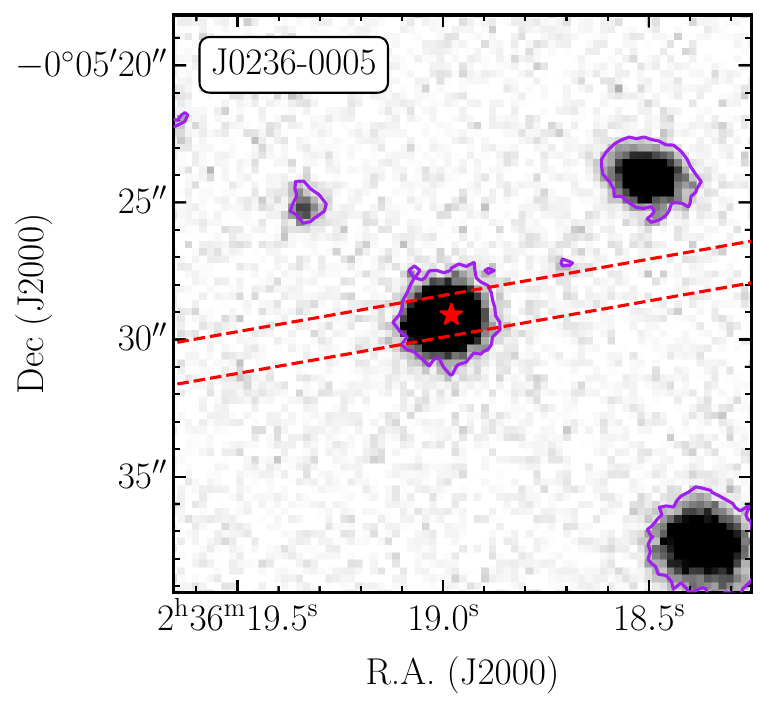}
  \end{subfigure}
  \begin{subfigure}{0.19\textwidth}
    \centering\includegraphics[width=\textwidth]{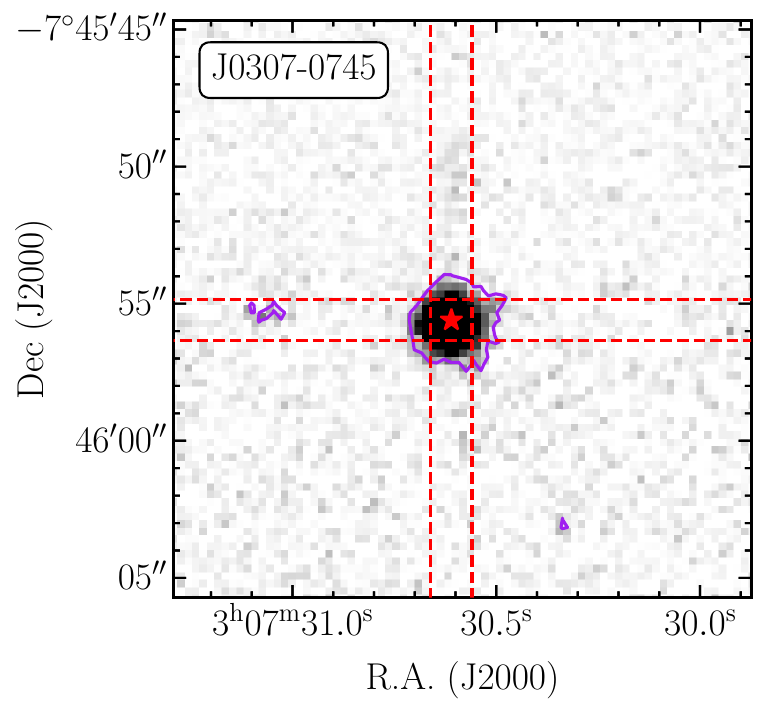}
  \end{subfigure}
  
  \begin{subfigure}{0.19\textwidth}
    \centering\includegraphics[width=\textwidth]{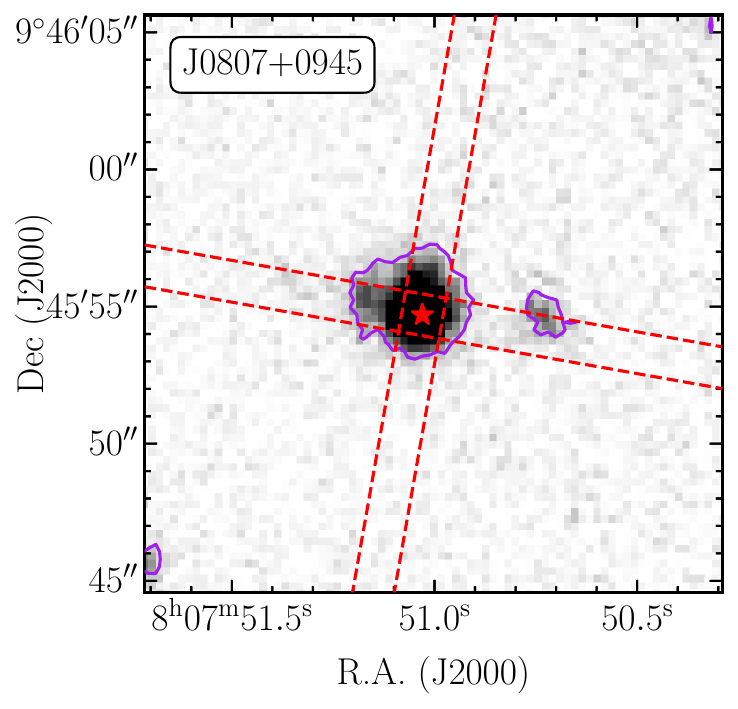}
  \end{subfigure}
  \begin{subfigure}{0.19\textwidth}
    \centering\includegraphics[width=\textwidth]{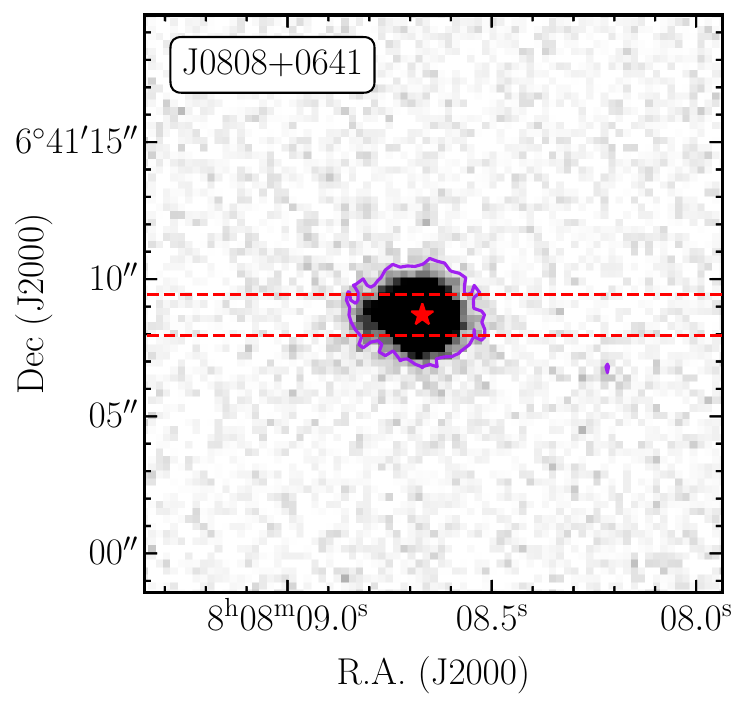}
  \end{subfigure}
  \begin{subfigure}{0.19\textwidth}
    \centering\includegraphics[width=\textwidth]{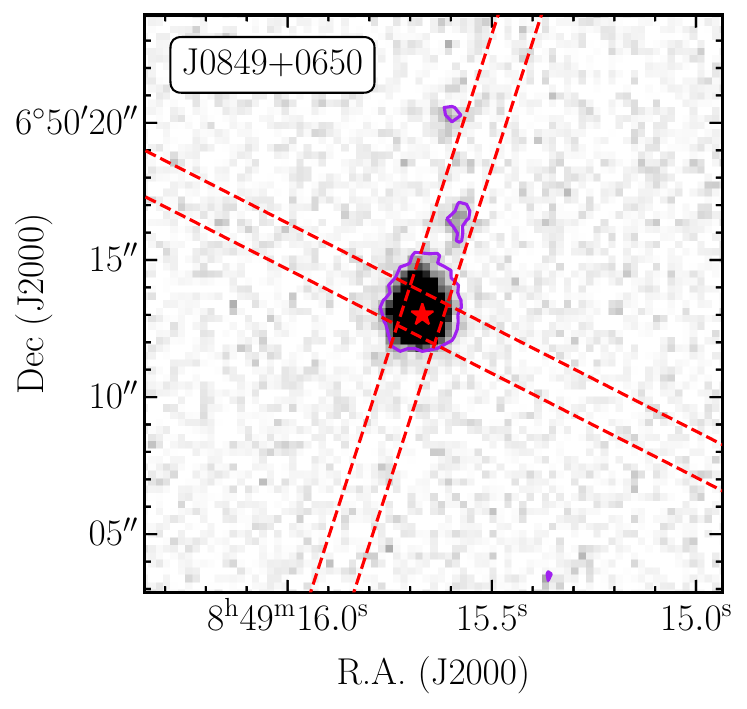}
  \end{subfigure}
  \begin{subfigure}{0.19\textwidth}
    \centering\includegraphics[width=\textwidth]{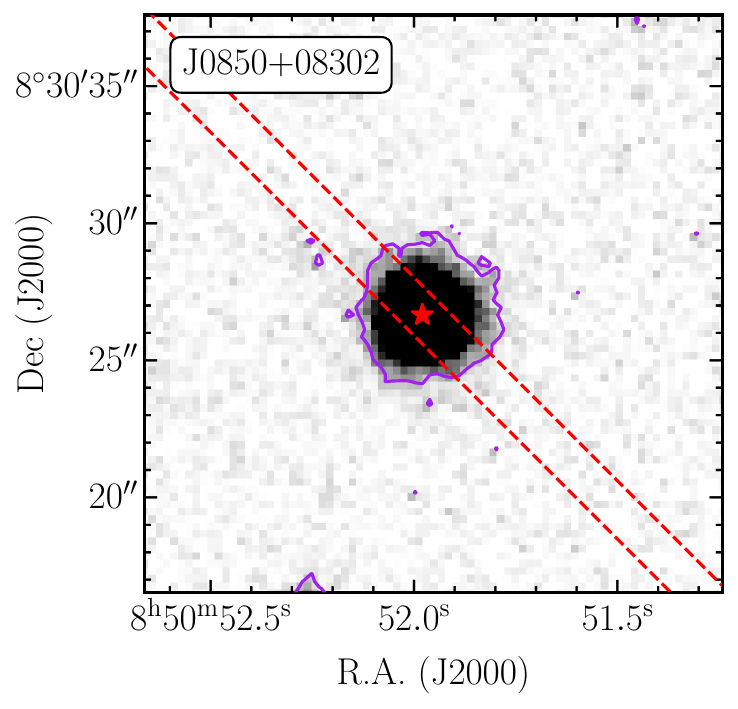}
  \end{subfigure}
  \begin{subfigure}{0.19\textwidth}
    \centering\includegraphics[width=\textwidth]{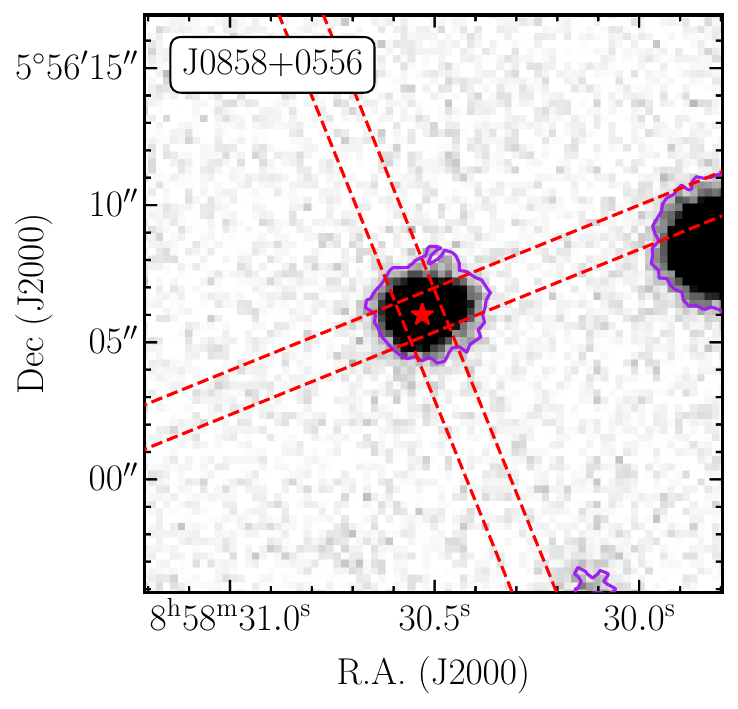}
  \end{subfigure}
  
  \begin{subfigure}{0.19\textwidth}
    \centering\includegraphics[width=\textwidth]{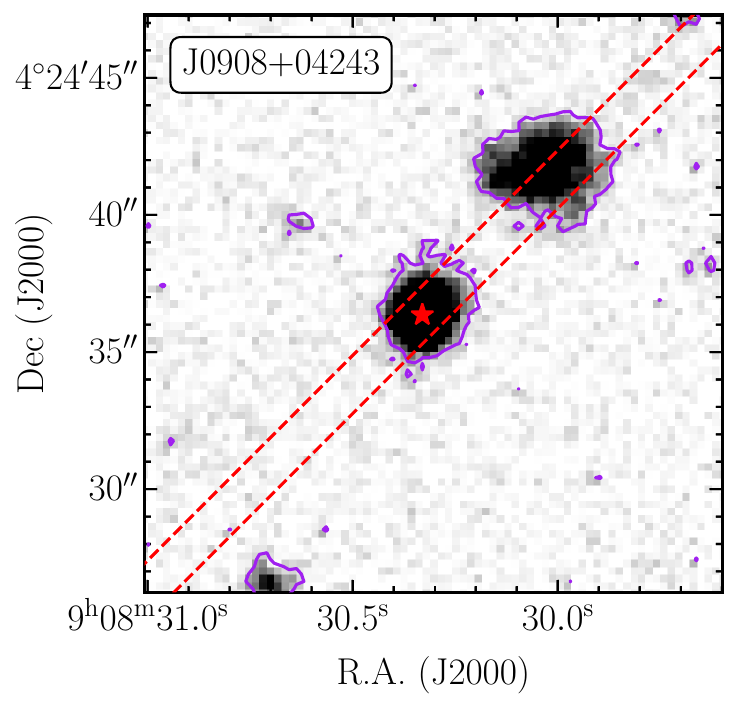}
  \end{subfigure}
  \begin{subfigure}{0.19\textwidth}
    \centering\includegraphics[width=\textwidth]{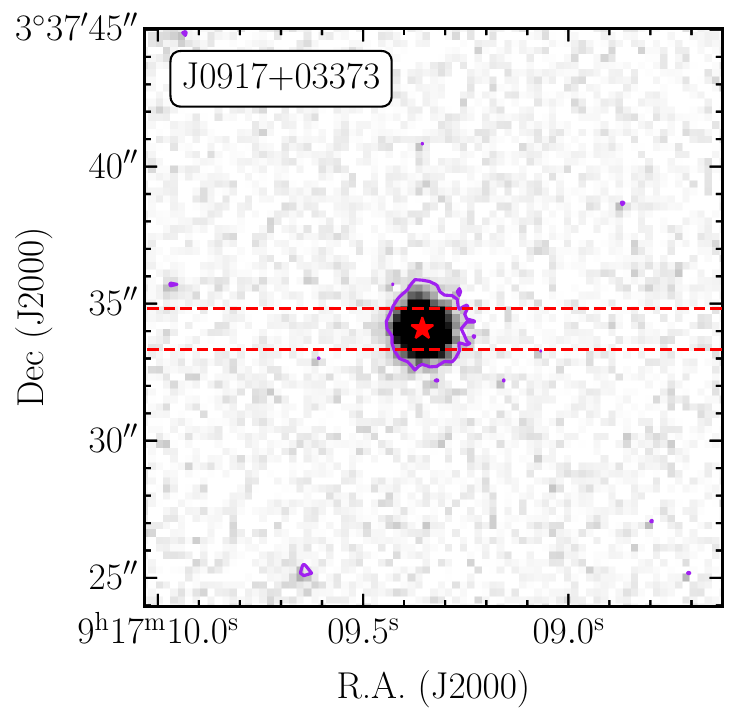}
  \end{subfigure}
  \begin{subfigure}{0.19\textwidth}
    \centering\includegraphics[width=\textwidth]{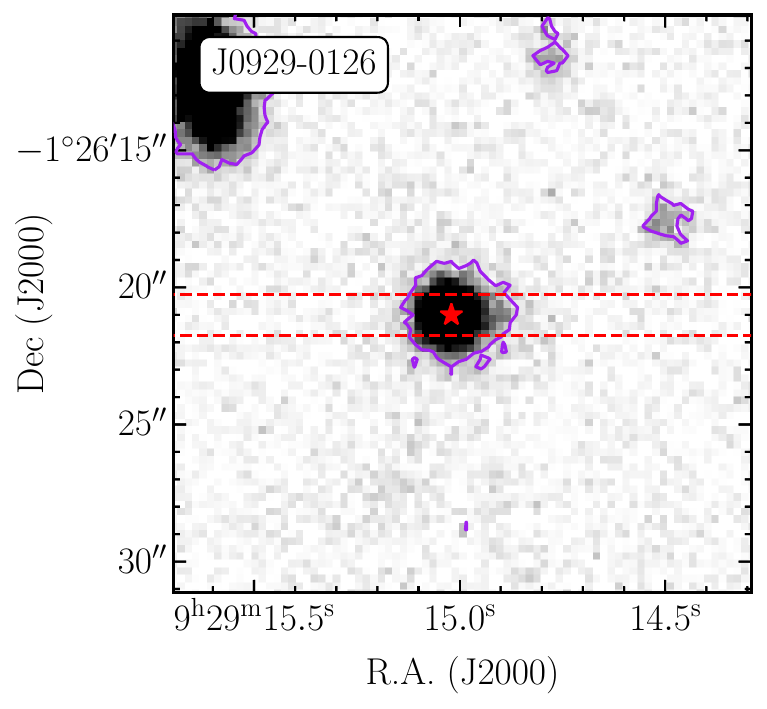}
  \end{subfigure}
  \begin{subfigure}{0.19\textwidth}
    \centering\includegraphics[width=\textwidth]{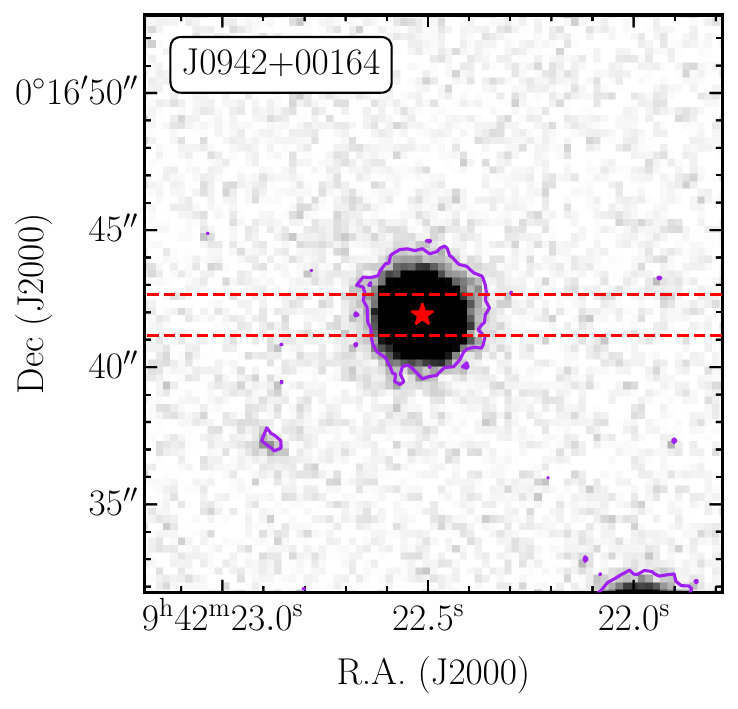}
  \end{subfigure}
  \begin{subfigure}{0.19\textwidth}
    \centering\includegraphics[width=\textwidth]{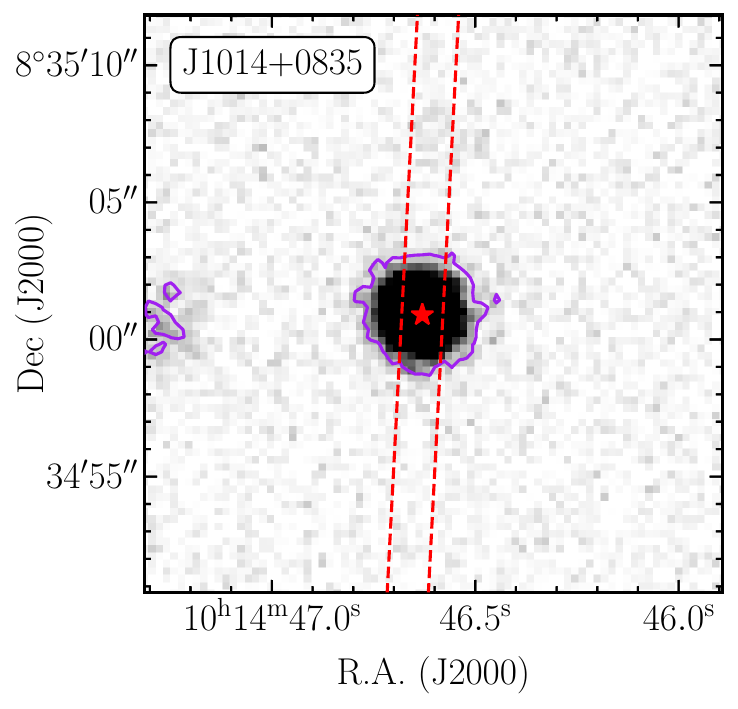}
  \end{subfigure}
  
  \begin{subfigure}{0.19\textwidth}
    \centering\includegraphics[width=\textwidth]{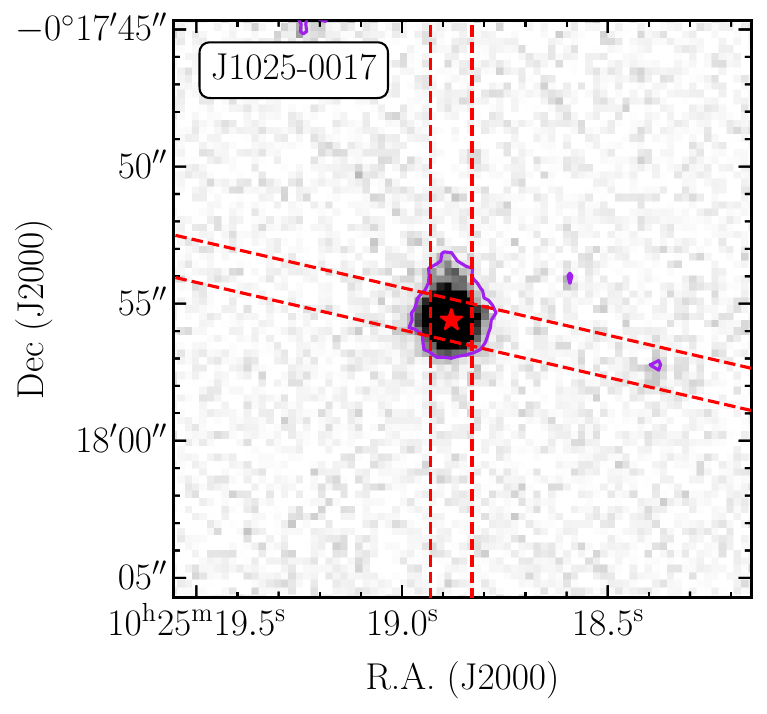}
  \end{subfigure}
  \begin{subfigure}{0.19\textwidth}
    \centering\includegraphics[width=\textwidth]{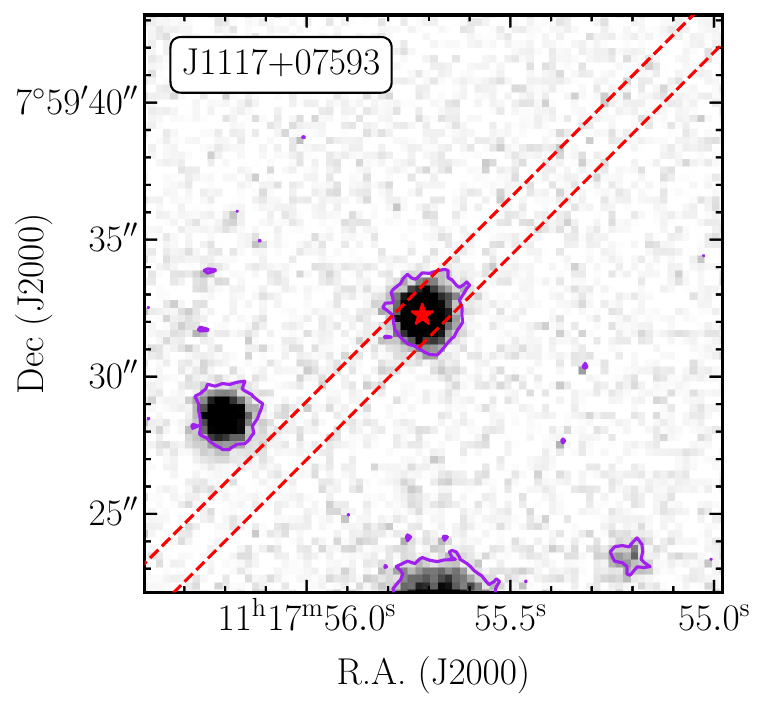}
  \end{subfigure}
    \begin{subfigure}{0.19\textwidth}
    \centering\includegraphics[width=\textwidth]{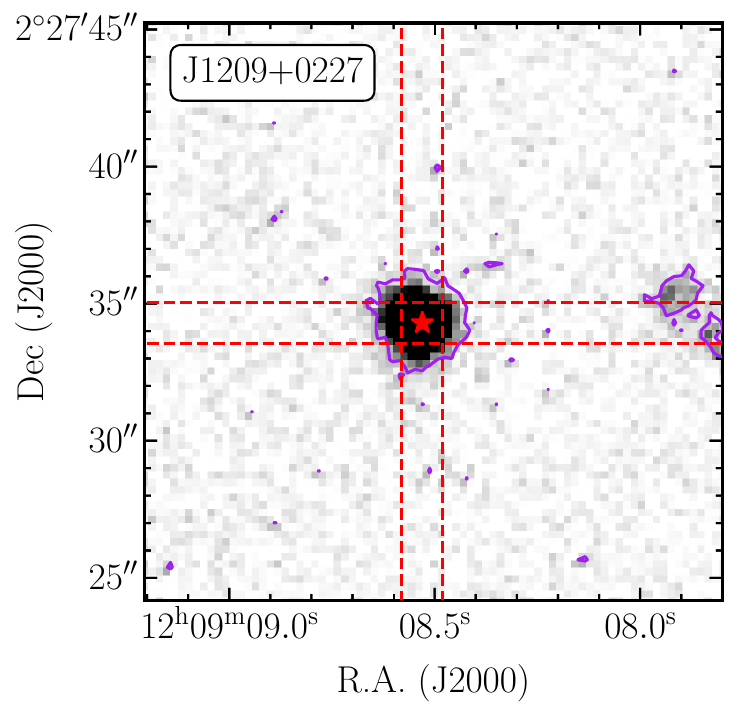}
  \end{subfigure}
  \begin{subfigure}{0.19\textwidth}
    \centering\includegraphics[width=\textwidth]{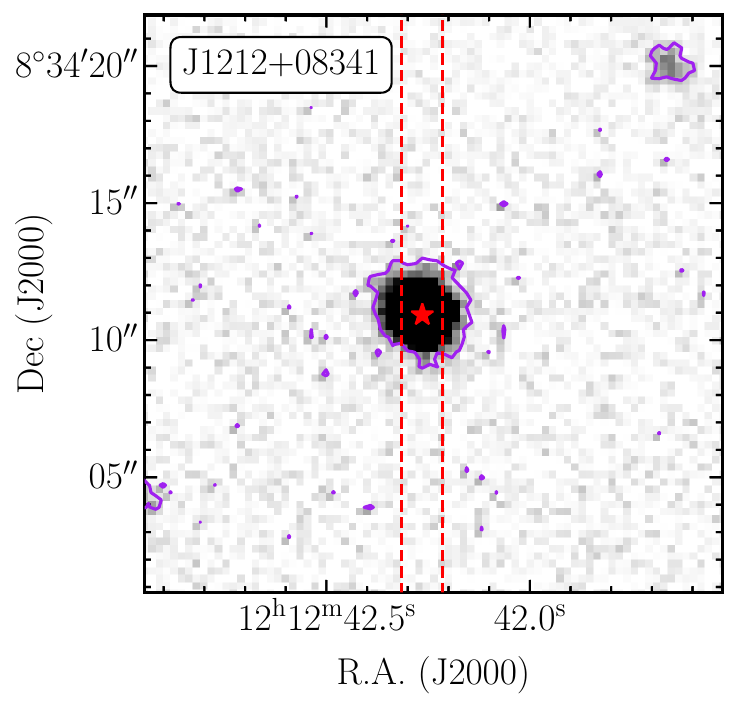}
  \end{subfigure}
  \begin{subfigure}{0.19\textwidth}
    \centering\includegraphics[width=\textwidth]{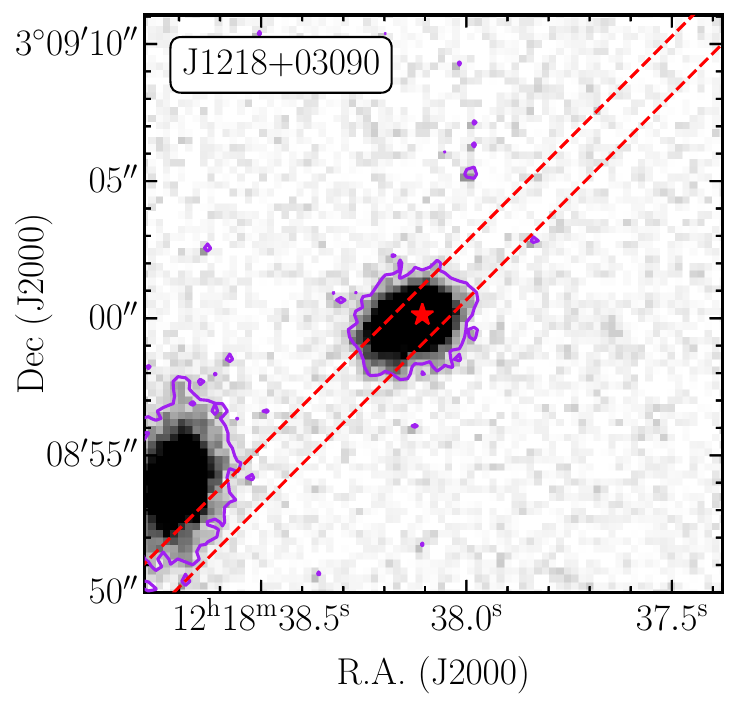}
  \end{subfigure}
  
  \begin{subfigure}{0.19\textwidth}
    \centering\includegraphics[width=\textwidth]{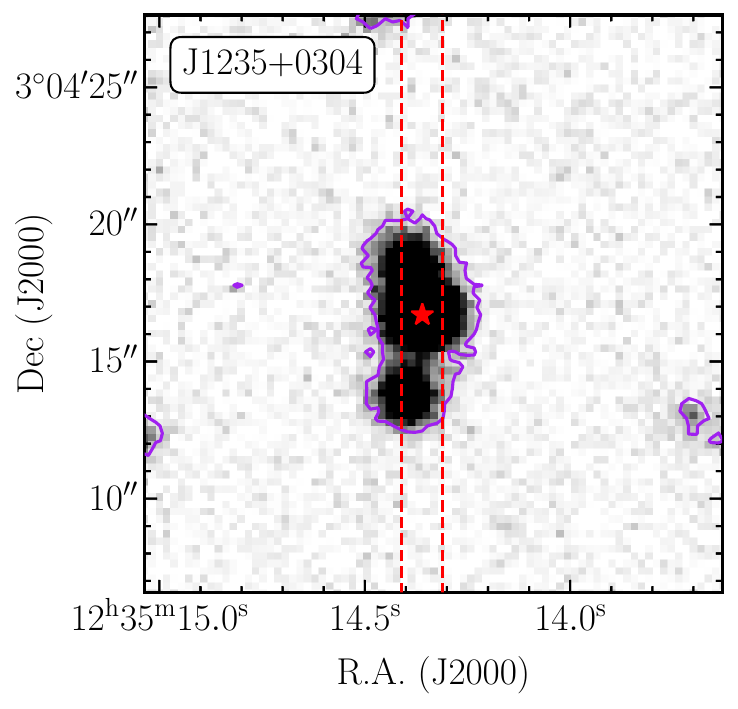}
  \end{subfigure} 
  \begin{subfigure}{0.19\textwidth}
    \centering\includegraphics[width=\textwidth]{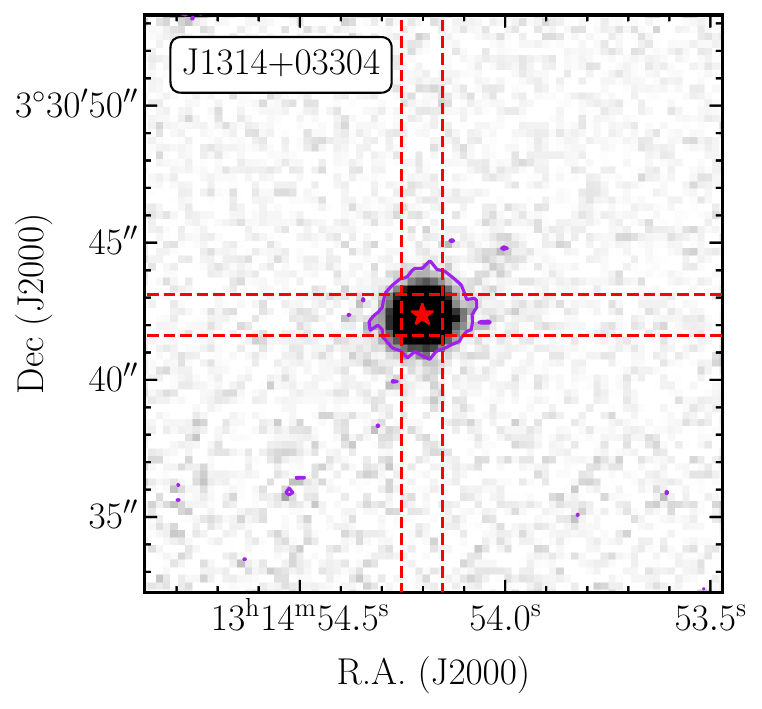}
  \end{subfigure}
  \begin{subfigure}{0.19\textwidth}
    \centering\includegraphics[width=\textwidth]{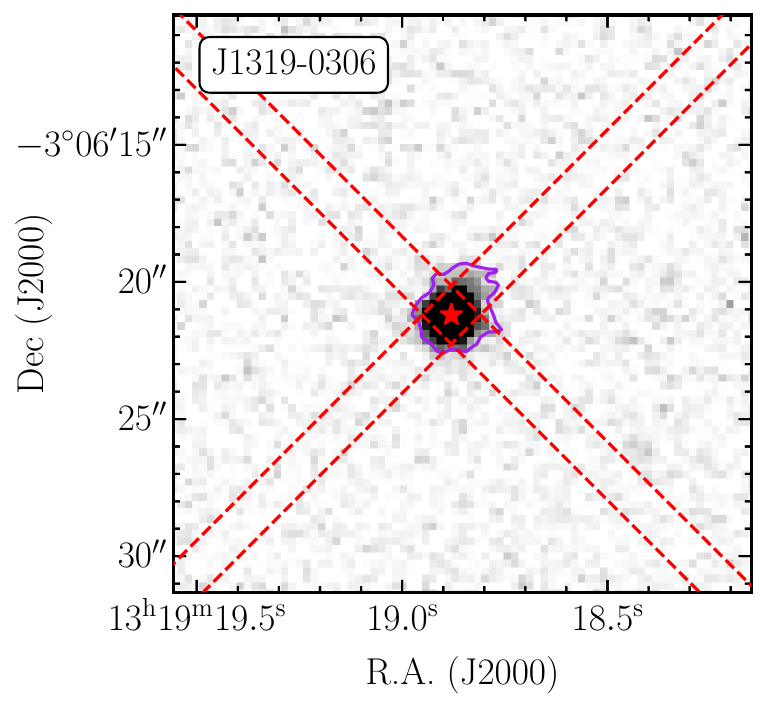}
  \end{subfigure}
  \begin{subfigure}{0.19\textwidth}
    \centering\includegraphics[width=\textwidth]{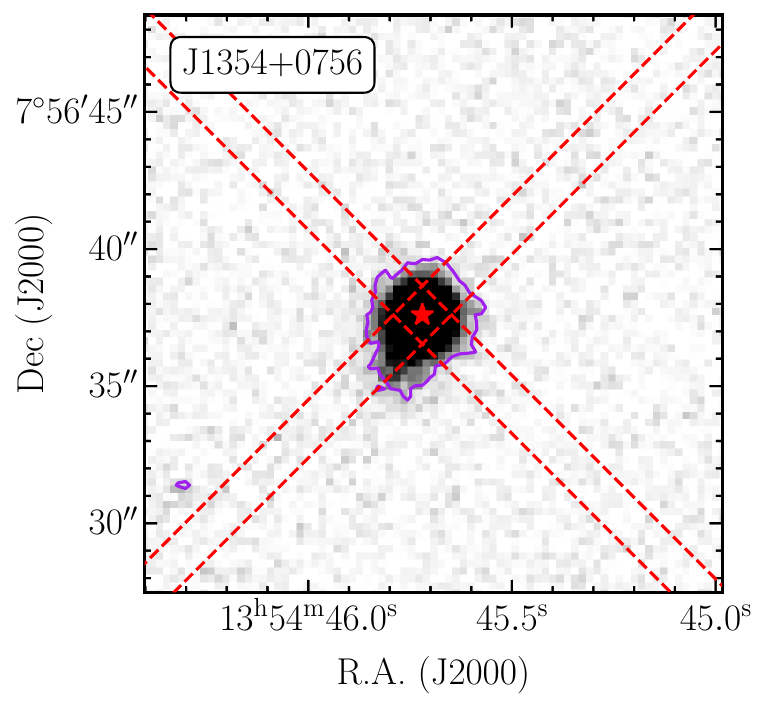}
  \end{subfigure}
  \begin{subfigure}{0.19\textwidth}
    \centering\includegraphics[width=\textwidth]{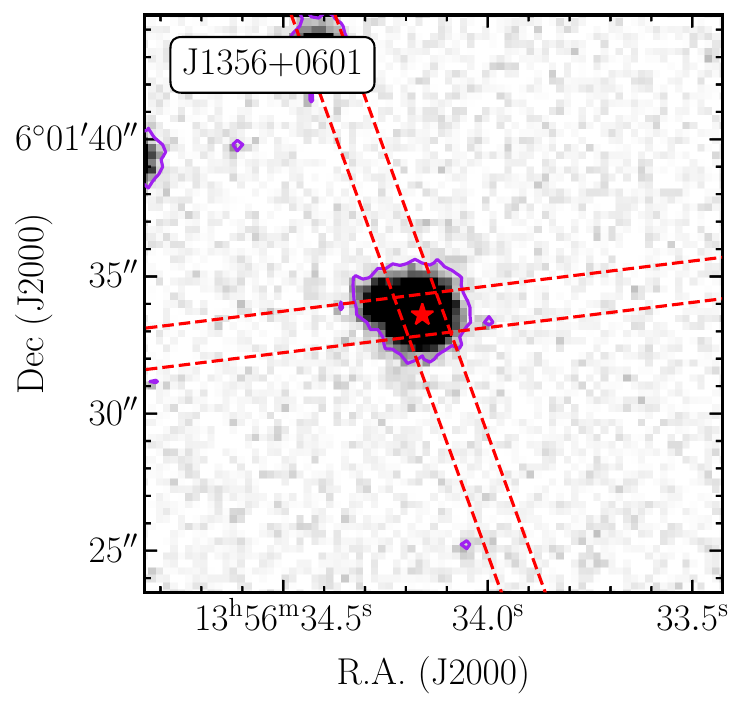}
  \end{subfigure}
  
  \begin{subfigure}{0.19\textwidth}
   \centering\includegraphics[width=\textwidth]{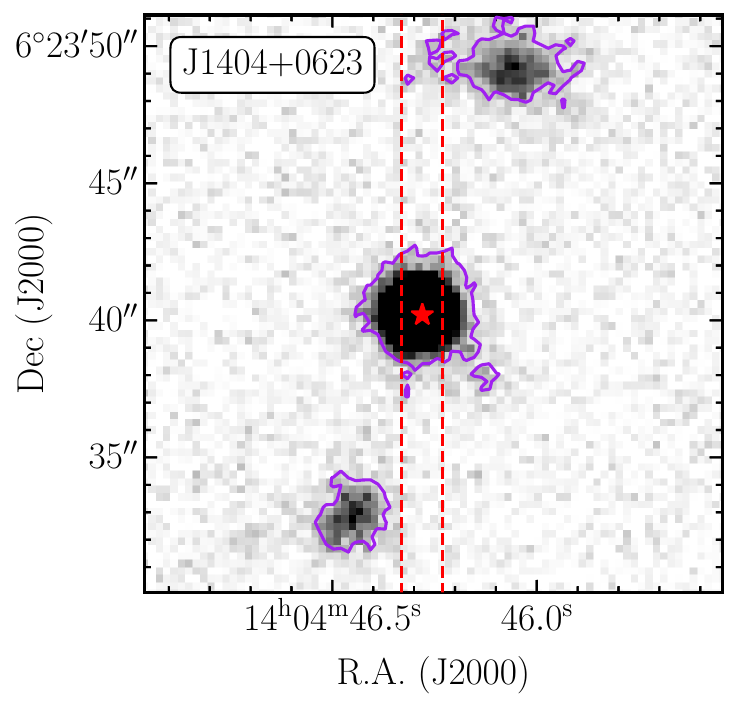}
  \end{subfigure}
  \begin{subfigure}{0.19\textwidth}
   \centering\includegraphics[width=\textwidth]{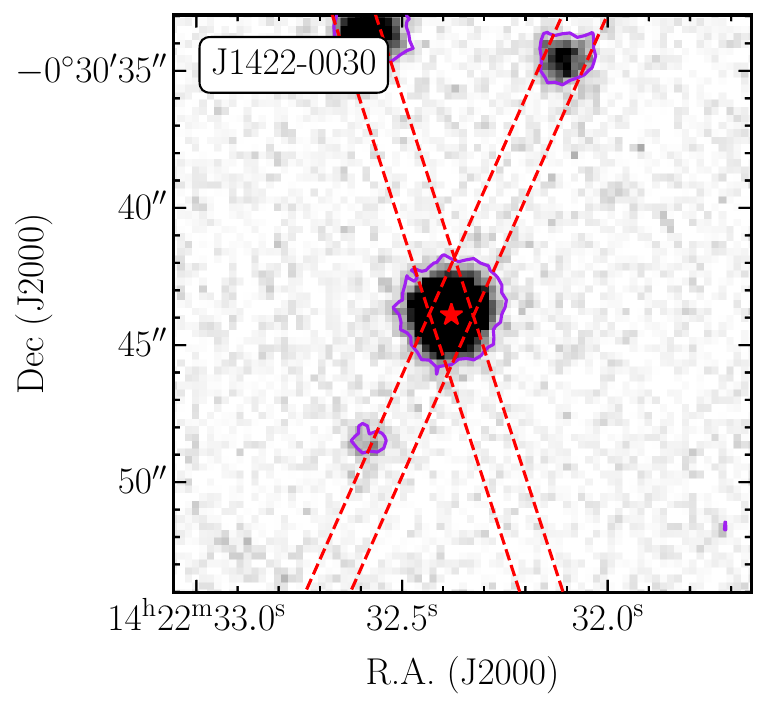}
  \end{subfigure}
  \begin{subfigure}{0.19\textwidth}
    \centering\includegraphics[width=\textwidth]{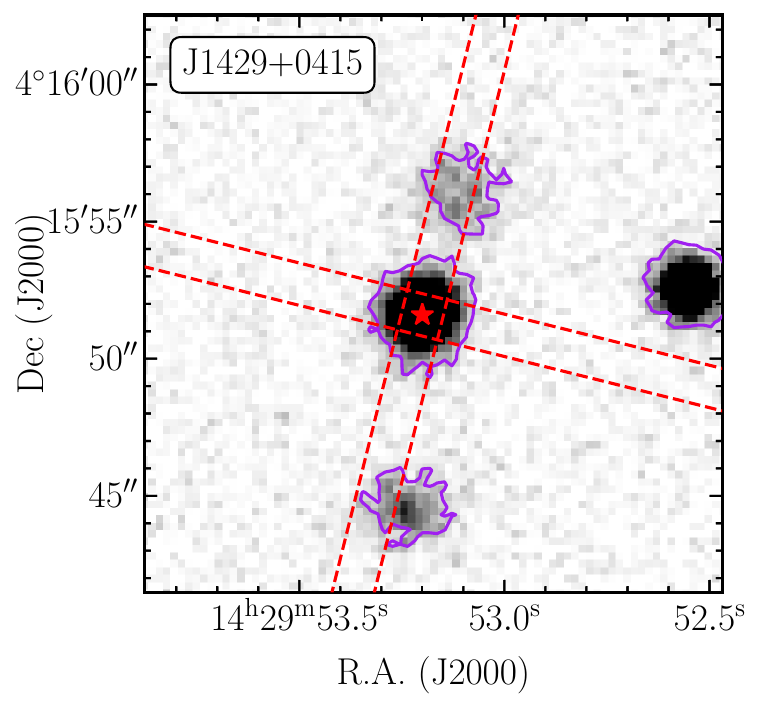}
  \end{subfigure}
  \begin{subfigure}{0.19\textwidth}
    \centering\includegraphics[width=\textwidth]{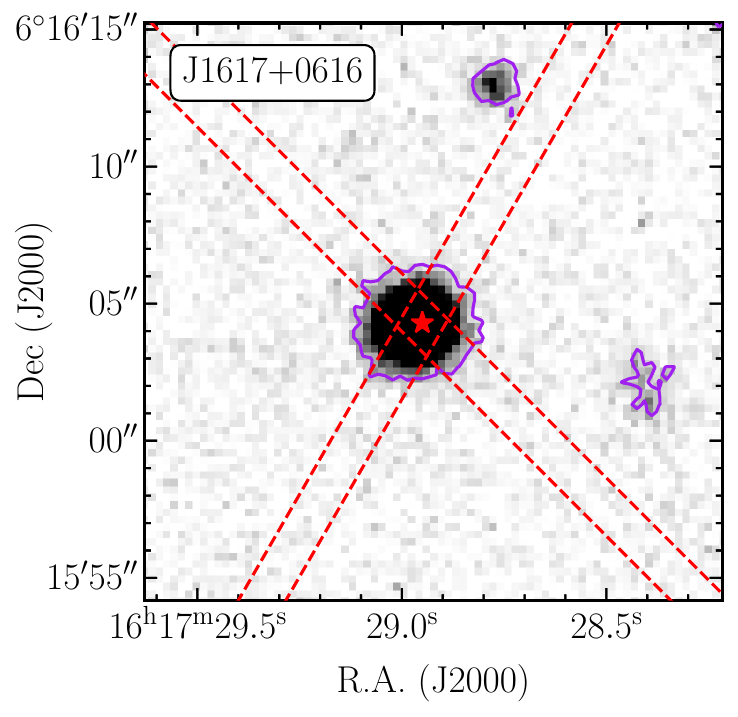}
  \end{subfigure}
      \begin{subfigure}{0.19\textwidth}
   \centering\includegraphics[width=\textwidth]{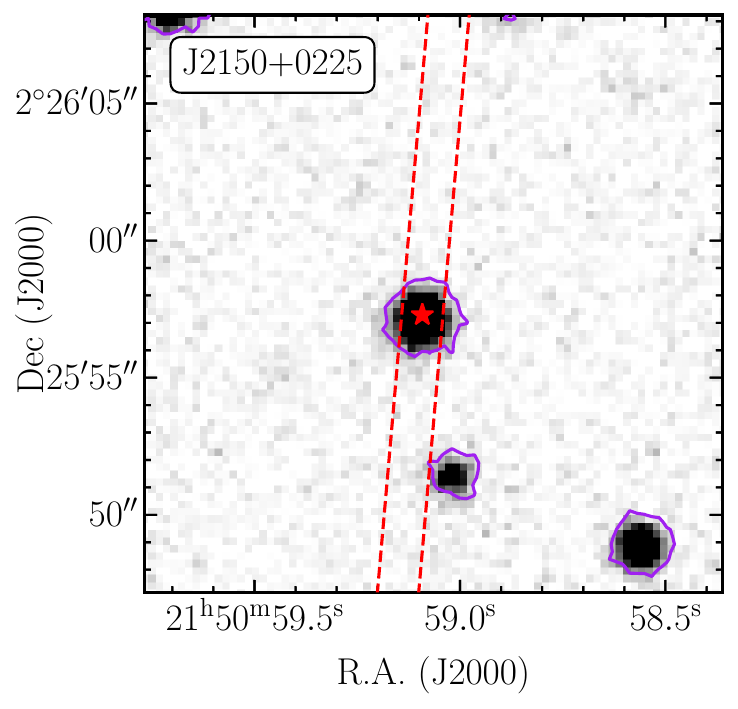}
  \end{subfigure}

  \caption{Slit configurations used in SALT to observe the GOTOQs to calculate the impact parameters precisely are shown on top of the DECaLS r-band images. The quasar sits at the center of each image and is marked with red stars. Dotted horizontal lines indicate the long-slit configuration used for our observations.}
  \label{fig:qso_fields}
\end{figure*}

\begin{figure*} 
  \begin{subfigure}{0.19\textwidth}
    \centering\includegraphics[width=\textwidth]{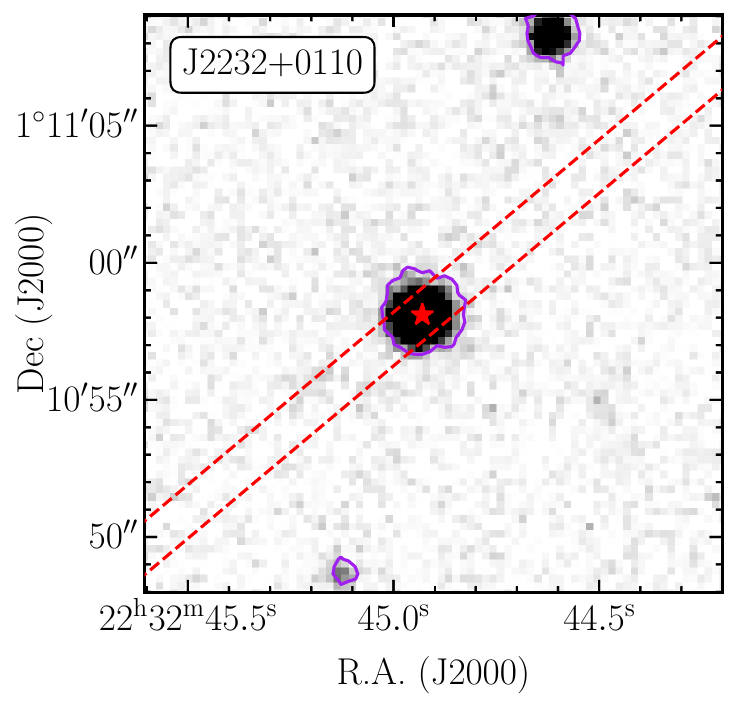}
  \end{subfigure}
   \begin{subfigure}{0.19\textwidth}
    \centering\includegraphics[width=\textwidth]{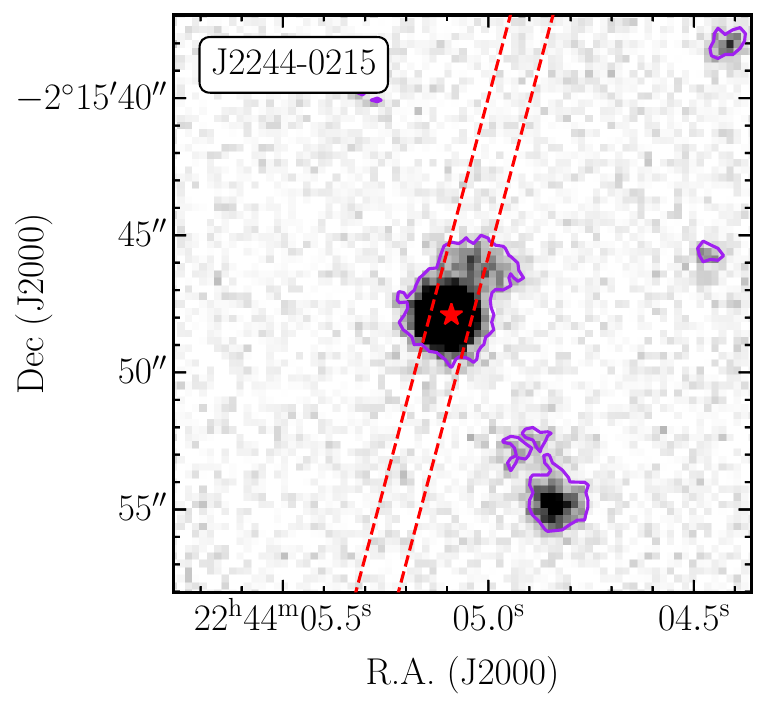}
  \end{subfigure}
    \begin{subfigure}{0.19\textwidth}
    \centering\includegraphics[width=\textwidth]{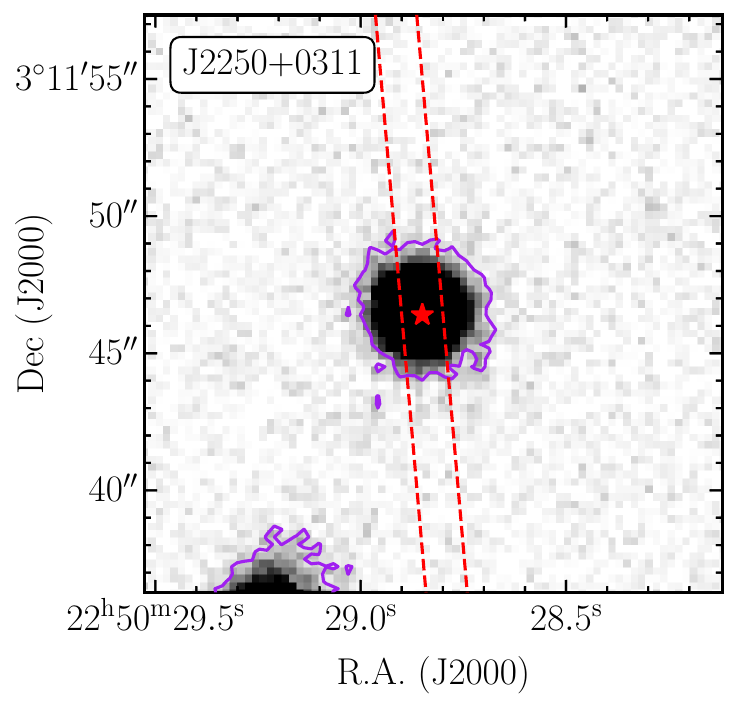}
  \end{subfigure}
  \begin{subfigure}{0.19\textwidth}
  \centering\includegraphics[width=\textwidth]{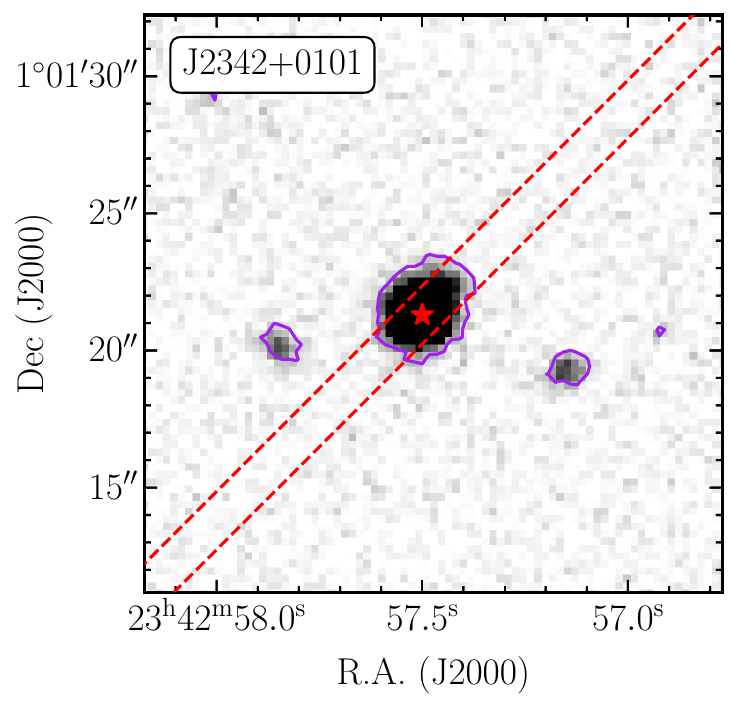}
  \end{subfigure}
  \begin{subfigure}{0.19\textwidth}
    \centering\includegraphics[width=\textwidth]{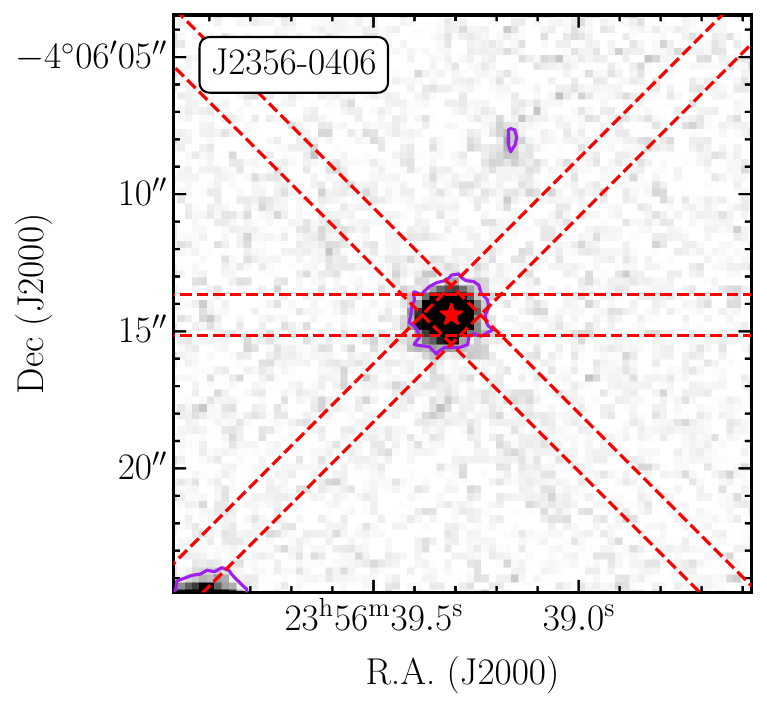}
  \end{subfigure}
  \setcounter{figure}{2}
  \caption{Continued.}
\end{figure*}

The details of the observations of the GOTOQs systems are given in Table~\ref{tab:observations}. 
The SALT spectroscopic observations were performed using the Robert Stobie Spectrograph (RSS) in long-slit mode from April 2017 to { December 2023} (Program IDs: 2017-1-SCI-011, 2021-2-SCI-012, 2022-1-SCI-016, 2022-2-SCI-015, 2023-1-SCI-014, { 2023-2-SCI-004}). For our observations, we use the PG0900 grating along with a long slit-width of width $1.5^{\prime\prime}$ and the grating angles are so chosen that the expected nebular lines from the GOTOQ host galaxies fall within the wavelength coverage of the spectrograph avoiding the CCD gaps.

The slit configurations used are shown in Figure \ref{fig:qso_fields} on top of the Dark Energy Camera Legacy Survey (DECaLS) r-band images of the quasar fields. { SALT spectra were obtained for two or more PAs for 19 cases. In the remaining 21 cases we got spectra for one PA only. In ten of these sources the slit was aligned along the extensions seen in the images due to the presence of the foreground galaxies. For 11 sources (where there is no asymmetry clearly visible in the quasar images) the spectroscopic observations were not completed along the second PA. }


The raw CCD frames obtained from the SALT observations were initially subjected to the SALT data reduction and analysis pipeline \citep{crawford2010}. We then used the standard \texttt{IRAF} \citep{tody1986} routines to obtain the wavelength-calibrated 2D spectra and the flux-calibrated 1D spectra of the quasars. In summary, each science frame was first flat-field corrected, cosmic ray zapped, and then wavelength calibrated against a standard lamp. Then we applied the extinction correction due to the Earth's atmosphere, and then the 1D spectra of the GOTOQs were extracted. The long-slit spectra were all individual flux calibrated against standard stars observed with the same settings as the GOTOQs. Once the flux calibration is done, we apply the air to vacuum wavelength transformation and also correct for the heliocentric velocity.

{ Our SALT observations reconfirm the presence of nebular \OII\ emission from foreground galaxies for 39 out of 40 systems observed. In the case of J1314+0330, we do not detect \OII\ emission. There are three SDSS spectra, obtained during three widely separated epochs,  available for this source. The \OII\ emission is seen in only one spectrum that too in the one obtained with the narrower fiber. Therefore, we believe the detection of \OII\ emission in this case is false.}

\section{Results}
\label{sec:results}

\subsection{Measuring the impact parameters}
As mentioned earlier, the primary goal of this study is to measure the impact parameters of the GOTOQs using long-slit spectroscopy. One is looking for the offset of \OII\ emission with respect to the quasar trace at different position angles (PAs). We use these off-sets measured at different PAs to find the centroid of the \OII\ emission and hence the galaxy impact parameter (denoted as ${\rm D_{sp}})$. Typically, one requires three PAs to have much better centroiding of the galaxy. However, due to observing time constraints, we could not get this for all the objects in our sample. However, we address this shortcoming by measuring the impact parameter (i.e., denoted by ${\rm D_{ph}}$) using the residual image after subtracting the quasar contribution whenever possible. 
\begin{figure*}
\begin{minipage}{0.495\linewidth}
\centering
\includegraphics[width=0.995\textwidth]{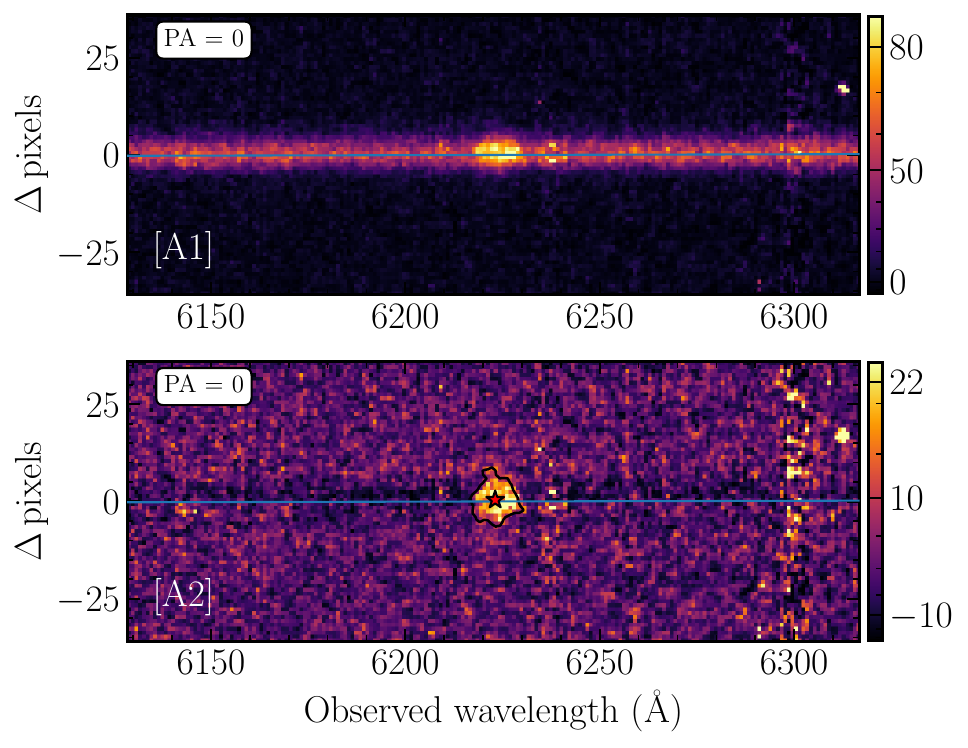}
\end{minipage}
\begin{minipage}{0.495\linewidth}
\centering
\includegraphics[width=0.995\textwidth]{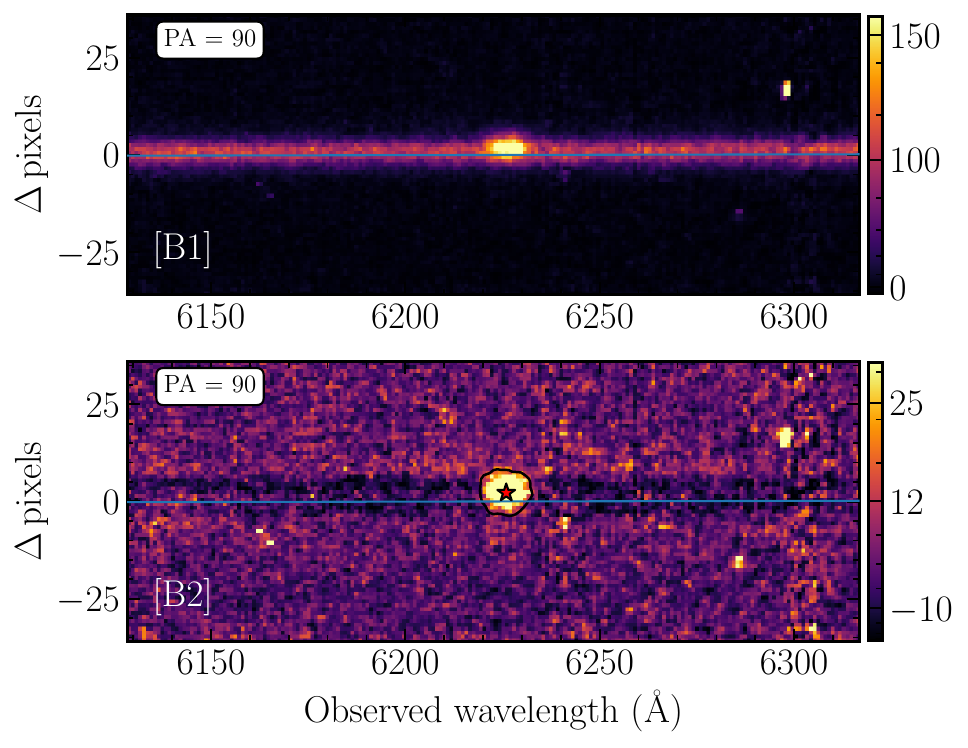}
\end{minipage}
\begin{minipage}{0.495\linewidth}
\centering
\includegraphics[width=0.95\textwidth]{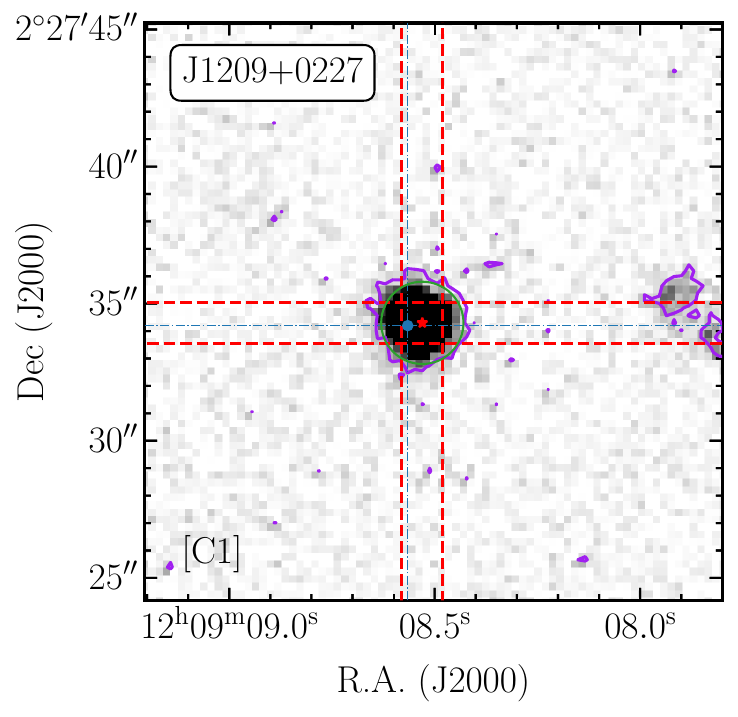}
\end{minipage}
\begin{minipage}{0.495\linewidth}
\centering
\includegraphics[width=0.95\textwidth]{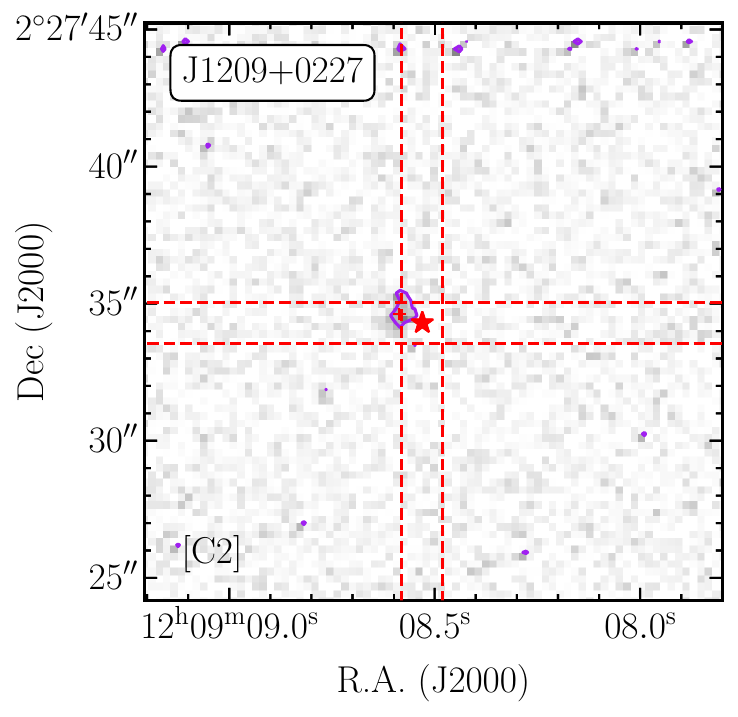}
\end{minipage}
\caption{{{The process of determining the \OII\ emission centroids for the GOTOQ system J1209+0227. Panel A1 shows a cutout of the 2D SALT spectrum for the position angle of 0$^\circ$ containing the quasar trace as well as the \OII\ emissions from the foreground galaxy color-coded according to the counts in each pixel. The solid blue line corresponds to the center of the quasar trace. The 2D spectrum upon removing the quasar trace is shown in the panel A2. The black contour around the \OII\ emission blob corresponds to the 3$\sigma$ contour above the average background. The red star marks the centroid of this 3$\sigma$ contour. Similarly, the B1 and B2 right panels show the process of identifying the centroid of the \OII\ emission for the position angle of 90$^\circ$. Panel C1 shows the DECaLS r-band image of the quasar field. The red dashed vertical and horizontal pair of lines corresponding to the slit configuration for the position angles of 0$^\circ$ and 90$^\circ$ overlayed on the image. The red star marks the center of the quasar. Blue vertical and horizontal dot-dashed lines correspond to the offsets of the \OII\ emission centroid from the quasar for the slit position angles of 90$^\circ$ and 0$^\circ$. The intersection of these two lines is marked by a blue circle denoting the center of the foreground galaxy. The impact parameter is then given by the distance between the red star and the blue circle, which is, in this case, 3.8 kpc. The solid green circle corresponds to the SDSS fiber used to observe the quasar. Panel C2 shows the DECaLS r-band residual image and the excess residual flux is consistent with the location of the foreground galaxy measured spectroscopically. The red dashed pairs of lines correspond to the slit configurations.}}}
\label{fig:qso_subtract}
\end{figure*}

In the following, we illustrate the ${\rm D_{sp}}$ measurement for the GOTOQ towards J1209+0227. This is similar to what has been described in \citet{Noterdaeme_2010}. We begin with the reduced wavelength calibrated and background subtracted 2D SALT spectrum for the slit position angle of 0$^\circ$. We first model the quasar trace by a Gaussian in the spatial direction, whose amplitude is allowed to vary smoothly (a third-order polynomial) along the wavelength axis. The centroid of the Gaussian is allowed to vary linearly with the wavelength to account for the possible misalignment between the quasar trace and the pixels of the CCD. Then we subtract this model from the original 2D SALT spectrum to obtain the \OII\ emission without the quasar contamination as much as possible. We identify the centroid of this emission to measure the offset of the \OII\ emission relative to the background quasar trace. This process is illustrated in the top and middle left panels ({ panel [A1] and [A2]}) of Figure \ref{fig:qso_subtract}. The solid blue line corresponds to the quasar trace, while the red star marks the centroid of the \OII\ emission. The centroid is found to lie 0.4 pixels above the central quasar trace. We perform the same exercise for the spectra obtained at a position angle of 90$^\circ$, illustrated in the top and middle right panels ({ panel [B1] and [B2]}) of Figure \ref{fig:qso_subtract}, and the centroid is located 2.2 pixels above the central quasar trace. Once the centroids of the \OII\ emission are known along two different slit orientations, it is straightforward to measure the impact parameter of the GOTOQ. Since the background quasar and foreground galaxies are at very small angular separation, we use the plane geometry to find the impact parameter. Figure \ref{fig:qso_subtract} illustrates this procedure in the bottom panels ({ panel [C1] and [C2]}). For this case we get ${\rm D_{sp} = 3.8}$ kpc. For each GOTOQ, we also looked at the residual r-band image obtained after subtracting the Quasar PSF \footnote{Detailed procedure for obtaining the PSF is given in https://psfex.readthedocs.io/en/latest/Working.html} (as shown in the panel [C2] of Figure~\ref{fig:qso_subtract} for J1209+0227). We obtain ${\rm D_{ph} = 5.9}$ kpc. The residual image also demonstrates the need for the spectra at the third PA to get the accurate impact parameters. We use the ${\rm D_{phot}}$ for statistical analysis in such cases. We similarly calculate the impact parameters for the rest of the GOTOQs in our sample. { Note that when the residual flux is present after subtracting the quasar model, we attribute this residual flux to the foreground galaxy.}

\begin{table}
    \centering
    \begin{tabular}{>{\rowmac}l>{\rowmac}c>{\rowmac}c>{\rowmac}c<{\clearrow}}
     \hline
       No. & System & $\rm D_{sp}$ & $\rm D_{ph}$\\
            &        &  (kpc) &  (kpc) \\
        (1) & (2)    &  (3)    & (4)   \\
        \hline
        \hline
        
        1   &   J000123.61$-$032317.9 & --   &  --   \\
        2   &   J005408.46$-$094638.2 & 3.3  &  4.8  \\
        3   &   J005436.35+005055.6   & --   &  5.2  \\
        4   &   J012916.14$-$003211.4 & 13.6 &  13.5 \\
        5   &   J013132.92$-$040137.5 & --   &  10.3 \\
        6   &   J020317.22$-$010122.8 & 7.3  &  --   \\
        7   &   J023529.82+005916.3   & 9.3  &  6.3  \\
        8   &   J023608.18+005604.0   & 7.5  &  6.3  \\
        9   &   J023618.98$-$000529.1 & --   &  8.6  \\
        10  &   J030730.61$-$074555.6 & 1.2  &  7.4  \\
        11  &   J080751.03+094554.7   & 15.2 &  16.3 \\
        12  &   J080808.67+064108.7   & 8.6  &  9.2  \\
        13  &   J084915.67+065013.0   & 10.5 &  9.5  \\
        14  &   { J085051.97+083026.6}   & --   &   --  \\
        15  &   J085830.53+055606.0   & 7.0  &  7.7  \\
        16  &   { J090830.32+042436.3}   & --   &  --   \\
        17  &   { J091709.35+033734.0}   & --   &  --   \\
        18  &   J092915.02$-$012621.0 & 14.0 &  13.4 \\
        19  &   { J094222.51+001641.9}   & --   &  --   \\
        20  &   J101446.63+083500.9   & --   &  6.4  \\
        21  &   J102518.88$-$001755.6 & 10.0 &  11.2 \\
        22  &   { J111755.71+075932.2}   & --   &  8.4  \\
        23  &   J120908.53+022734.3   & 3.8  &  5.9  \\
        24  &   { J121242.26+083410.9}   & --   &  --   \\
        25  &   { J121838.10+030900.1}   & 9.3 &  11.4  \\
        26  &   J123514.36+030416.7   & --   &  --   \\
        27  &   J131918.88$-$030621.2 & --   &  7.8  \\
        28  &   J135445.72+075637.6   & 15.3 &  14.2 \\
        29  &   J135634.16+060133.6   & 10.1 &  11.2 \\
        30  &   J140446.28+062340.2   & --   &  3.6  \\
        31  &   J142232.38$-$003043.9 & 8.1  &  9.0  \\
        32  &   J142953.20+041551.6   & 13.9 &  --   \\
        33  &   J161728.95+061604.3   & 3.9  &  9.8  \\
        34  &   J215059.09+022557.2   &  --  &  5.7  \\
        35  &   J223244.93+011058.1   &  --  &  4.4  \\
        36  &   J224405.09$-$021547.9 & 16.3 &  16.3 \\
        37  &   J225028.85+031146.4   & --   &   --  \\
        38  &   J234257.50+010121.3   & 8.9  &   9.5 \\
        39  &   J235639.31$-$040614.4 & 6.2  &   --  \\
        \hline
    \end{tabular}
    \caption{Measurements of impact parameter of the GOTOQ sub-sample from the SALT observations based on the triangulation as well as from the residual image from DESI-LIS.}
    \label{table:OII_emission}
\end{table}

\begin{figure}
    \centering
    \includegraphics[width=0.45\textwidth]{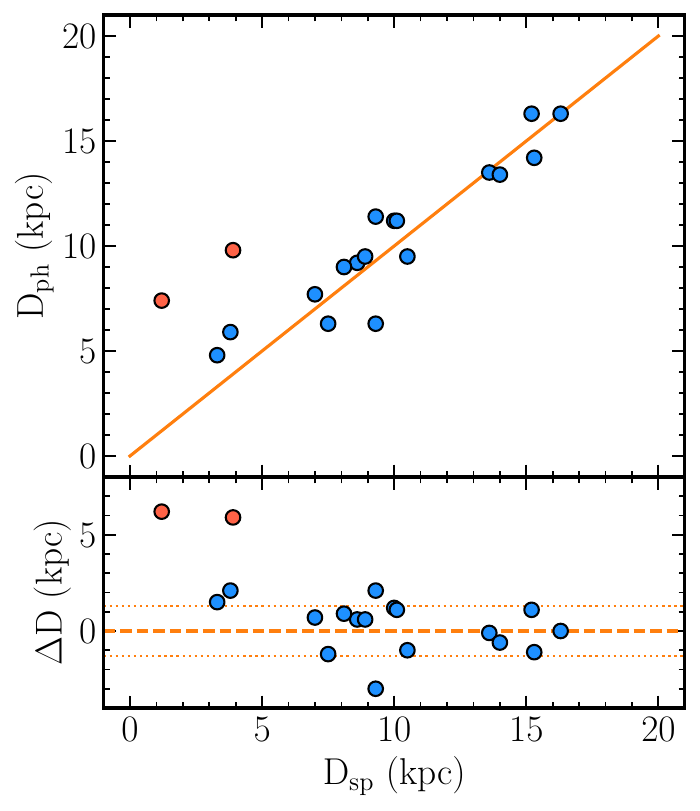}
    \caption{{{Top panel}}: Comparison between the impact parameters measured spectroscopically and photometrically. Solid orange line corresponds to $\rm D_{sp} = D_{ph}$.  {{Bottom panel}}: Residual impact parameter measured photometrically against the spectroscopically measured impact parameter. The orange dashed line shows $\rm D_{sp} = D_{ph}$. The dotted orange lines correspond to the standard deviation of the distribution. The red points correspond to the GOTOQs where large discrepancies between $\rm D_{ph}$ and $\rm D_{sp}$ are observed.}
    \label{fig:dsp_vs_dph}
\end{figure}

The impact parameters obtained for our sample are listed in the third and fourth columns of Table \ref{table:OII_emission}. 
{ We measure ${\rm D_{sp}}$ in the case of 22 GOTOQs. Based on residual broadband images and the photometric extensions, we measure ${\rm D_{phot}}$ for 28 GOTOQs. For 19 GOTOQs we have measured both ${ D_{sp}}$ and ${\rm D_{phot}}$. For 8 GOTOQs we could not measure either ${ D_{sp}}$ or ${\rm D_{phot}}$. Spectra of six of these eight GOTOQs were obtained along one PA only. We believe the impact parameters are small in these cases as nebular emission lines are aligned with the quasar trace. For the two cases, J0001$-$0323 and J1319$-$0306, we could not constrain the impact parameters of the foreground galaxies even with spectra along two different PAs. For J0001$-$0323, although we detect \OII\ emission from the foreground galaxy along PA=0 aligning with the quasar trace, we do not detect any emission from the galaxy along PA=150, implying the foreground galaxy to be present along the PA of 60. For J1319$-$0306, the SNR of the quasar trace at the 45-degree
position angle is too poor to determine the centroid of the \OII\ emission.}
We could measure the impact parameter for { 31} GOTOQs using spectroscopic or photometric methods. In the case of J1235+0304, as the QSO image is blended with two nearby sources, we used only one slit position angle for our spectroscopic observations covering the quasar and those two sources. We detect a broad \OII\ emission exactly on top of the quasar. Our analysis confirms that the two blended objects seen in the image are not at the redshift of the GOTOQs and, hence, are not the host galaxies of the \MgII\ absorption. No residual is detected when we model the image with three objects. In this case, we need spectra at another position angle to measure the impact parameter. Therefore, we do not include this object for further discussion. In three cases (GOTOQs towards J0203$-$0101, J2356$-$0406, and J1429+0415), we have measurements only based on long-slit spectroscopy as these systems have neither any photometric extension nor any residual left upon subtracting the quasar PSF. In the case of J2356$-$0406, the impact parameter measurement based on spectroscopy is secure as we use spectra obtained along three PAs.
{  Among the systems with $D_{phot}$ measured, for 12 cases, we have impact parameter measurements based on residual images rather than photometric extensions. However, in eight cases, we have spectroscopic confirmations of these residuals to be associated with foreground galaxies. For the remaining four cases, we have impact parameter measurements solely based on residual images.} 

For { 19} GOTOQs, we have impact measurements based on both photometry and spectroscopy. We compare the two measurements in Figure~\ref{fig:dsp_vs_dph}. Apart from two measurements (i.e., for J0307$-$0745 and J1617+0616 shown in red), other measurements are consistent within $\Delta D \lesssim 1.5$ kpc. In both these cases, the residual image is predominantly present in the regions not covered by our observations. We use the ${\rm D_{ph}}$ for all our analysis for these objects.

As expected, the measured impact parameters typically vary between 3 to 16 kpc with a median value of { 8.6 kpc}. For the median redshift (i.e., $z = 0.69$) of or sample 1" corresponds to 7.2 kpc. From Table~\ref{tab:sample}, we can find that { 19} GOTOQs in our sample are only detected from the BOSS spectra where the fiber used has a diameter of 2". In only { five} cases, the measured impact parameters are less than 7.2 kpc. Therefore, nearly all our measurements of \OII\ luminosity in the BOSS spectra are affected by fiber losses. On the other hand, { 10 out of 17} GOTOQs detected only in the SDSS spectra (with a fiber diameter of 3") have an impact parameter of less than 11 kpc. This implies that fiber loss effects may be less severe for these GOTOQs. There are { 21} measurements with impact parameters less than ten kpc. This is by far the largest sample of \MgII\ absorbers with spectroscopically confirmed host galaxies at such impact parameters. In what follows, we study the correlation between different absorption line parameters and impact parameters for this sample.

\subsection{Rest equivalent widths versus the impact parameters}

\begin{figure*}
    \centering
    \includegraphics[width=0.97\textwidth]{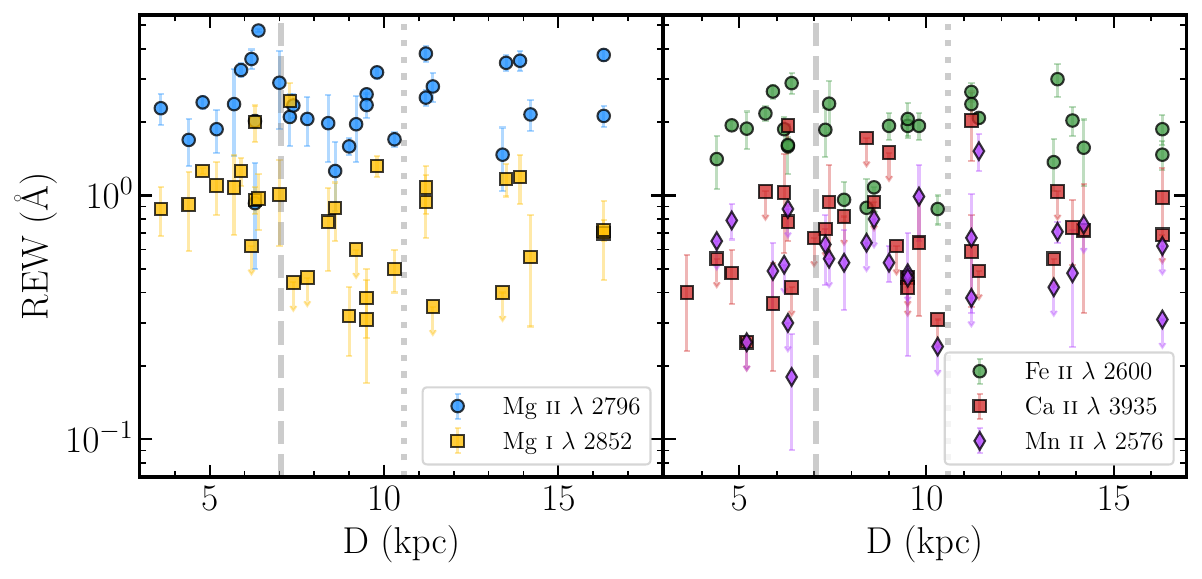}
    \caption{Rest equivalent widths of various neutral/singly ionized species versus the impact parameter. The left panel shows \MgII\ $\lambda$ 2796 line represented by blue circles and the \MgI\ $\lambda$ 2852 line depicted as yellow squares. The right panel shows \FeII\ $\lambda$ 2600 line (green circles), \CaII\ $\lambda$ 3935 line (red squares), and \MnII\ $\lambda$ 2576 (purple diamonds). The dashed and dotted vertical lines correspond to the physical angular scale of the SDSS DR12 and DR7 fibers at the median redshift of the observed GOTOQ subsample, respectively.}
    \label{fig:rew_vs_d}
\end{figure*}

In this section, we study the relationship between the rest equivalent width (REW) of different species and impact parameters. For this we consider Mg~{\sc ii} $\lambda$2796, Mg~{\sc i} $\lambda$2852, Fe~{\sc ii} $\lambda$2600, Ca~{\sc ii} $\lambda3935$ and Mn~{\sc ii} $\lambda2576$. As expected, we detect strong Fe~{\sc ii} and Mg~{\sc i} absorption with at least 2$\sigma$ significance in almost all cases. { We detect Ca~{\sc ii} $\lambda$3935 and Mn~{\sc ii} $\lambda$2576 absorption towards 16 and 17 GOTOQs in our observed sample, respectively.} In Figure~\ref{fig:rew_vs_d}, we plot the rest equivalent width of all the species as a function of the impact parameter. The figure also shows the impact parameter corresponding to 1" and 1.5" using long- and short-dashed vertical lines at the median redshift.  None of the species show any significant (anti-)correlation between the REW and the impact parameter.

\subsubsection{Ca~{\sc ii}  and Mg~{\sc ii} absorption:}
{ Earlier studies \citep{wild2006, sardane2015, Fang2023} suggested the existence of two populations of Ca~{\sc ii} absorbers defined based on the rest equivalent width of Ca~{\sc ii} (i.e., $W_{3935}$ above (strong) and below (weak) 0.7\AA). The strong Ca~{\sc ii} absorbers tend to produce stronger reddening signatures than weak Ca~{\sc ii} absorbers leading to the proposal that while strong Ca~{\sc ii} absorbers probe the inner dish+halo regions of galaxies, the weak Ca~{\sc ii} absorbers typically probe the halo (or CGM) of galaxies. One expects a possible anti-correlation between $W_{3935}$ and the impact parameter in such cases. In \citet{Guha2023}, using composite spectra, we showed the REW of Ca~{\sc ii} is higher for GOTOQs where the host galaxy is not well resolved compared to that of the resolved cases (i.e., their extended sample).}  

In our case, { 9 out of 16} detections have Ca~{\sc ii} equivalent width $\rm W_{3935}\geqslant 0.7$\AA. Thus, it belongs to the strong Ca~{\sc ii} absorbers population. This can be interpreted as the covering factor of strong Ca~{\sc ii} absorbers in the GOTOQs to be about { 23\%}. In Figure~\ref{fig:rew_vs_d}, we do not find any correlation between REW of Ca~{\sc ii} and $\rm D$. A good fraction of upper limits occurs at smaller impact parameters. Typically, they are around 0.6\AA\ due to poor SNR of the SDSS spectrum.  This is consistent with our finding that $\rm W_{2796}$ of the \usmg\ and DLAs do not correlate with the impact parameter. Lack of correlation between the rest equivalent width of Ca~{\sc ii} and D was also reported by \citet{Straka_2015} for their low-z GOTOQs. As in the case of Mg~{\sc ii}, one has to consider a larger impact parameter range to probe the dependence of Ca~{\sc ii} equivalent width and covering factor on the impact parameter.

\begin{figure}
    \centering
    \includegraphics[width=0.49\textwidth]{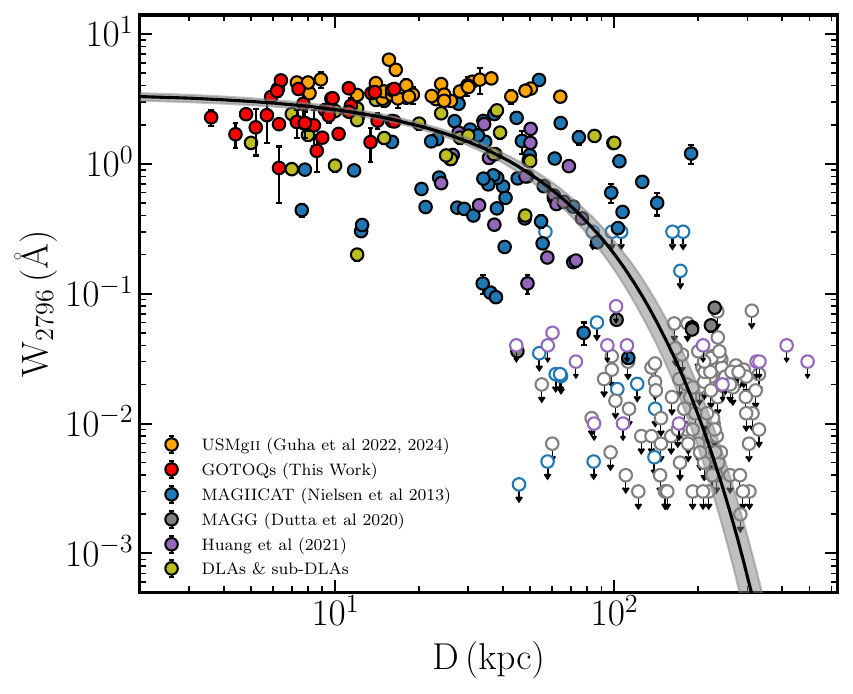}
    \caption{Impact Parameter versus the $\rm{W_{2796}}$ for this sample (red) along with studies from \citet{Guha2022, Guha2024} (orange), \citet{Nielsen_2013} (blue), \citet{Dutta2020} (grey) and, \citet{Huang2021} (purple) and sub-DLAs and DLAs ( green), taken from \citet{Rahmani2016}. Detected \MgII\ absorptions are marked by color-filled circles, whereas the upper limits are marked with hollow circles. The solid black line corresponds to the maximally likely log-linear fit to the observed data. The gray-filled region around the solid line corresponds to the $\rm{1\sigma}$ uncertainty to the fit.}
    \label{fig:w_vs_d}
\end{figure}

Note that about {25\%} of our systems have $\rm W_{2796}$ consistent with them being defined as \usmg. In the detailed analysis presented in \citet{Guha2022} and \citet{Guha2024}, we have shown that \usmg\ absorbers do not show any anti-correlation between $\rm W_{2796}$ and D. One of the possible reasons is that a good fraction of \usmg\ may originate from more than one galaxy. In the present case, it is most likely that the observed \MgII\ absorption is associated with a single galaxy. From Figure~\ref{fig:rew_vs_d}, we do not find any correlation between $\rm W_{2796}$ and D. It is also evident from the figure that the \usmg\ systems and the remaining ones are distributed uniformly across the impact parameter probed.

In \citet{Guha2023}, we have shown that including GOTOQs in the $\rm W_{2796}$ vs $\rm D$ is crucial to distinguish between a simple power-law fit and a log-linear fit. However, in that work, the impact parameters for the GOTOQs were obtained, assuming that the galaxy identified in the DECaLS photometry is the real host. Here, we repeat the exercise considering only the GOTOQs studied here, focusing mainly on \MgII\ absorbers at $z > 0.35$. 

To compare the GOTOQs studied in our sample with other \MgII\ absorption systems, we plot the same from the studies by \cite{Guha2022}, \citet{Guha2024}, \citet{Dutta2020, Huang2021, Nielsen_2013}. This is presented in Figure \ref{fig:w_vs_d}. The solid black line, which corresponds to the maximally likely log-linear fit to the combined observed data, is represented by the equation $\rm{{\log\, W_{2796} = (-0.013 \pm 0.002) D + (0.541 \pm 0.032)}}$ with an intrinsic scatter of ${\sigma_c = (1.213\pm0.075)}$ and the gray shaded region corresponds to the $1\sigma$ error associated with the fit. The fitting procedure, including the upper limits, is described in detail in the appendix of \citet{Guha2022} and is similar to what is described in \citet{Chen_2010, Dutta2020}. The fit values are consistent with those in \citet{Guha2024}.

\subsection{Line of sight reddening towards GOTOQ}
\begin{figure}
    \centering
    \includegraphics[width=0.45\textwidth]{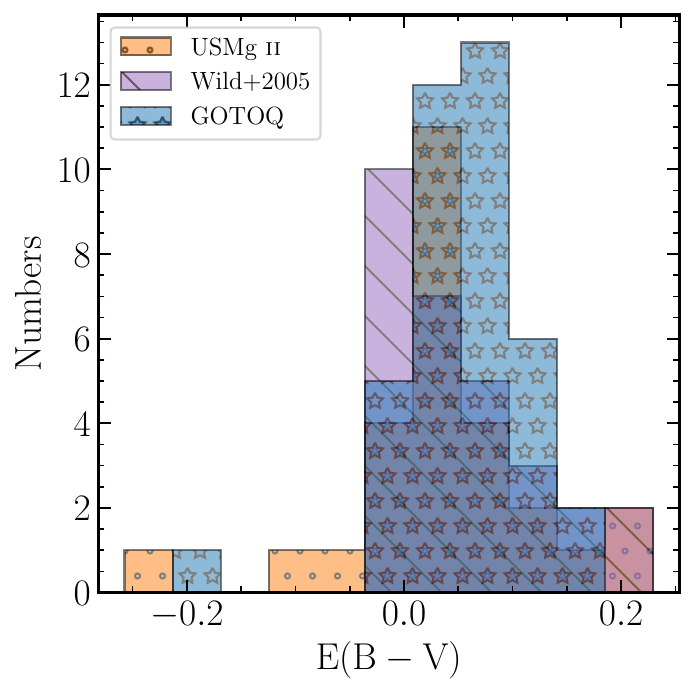}
    \caption{Comparison between the line of sight reddening towards the \usmg\ absorption systems (orange dotted histogram) at $z \sim 0.5$ \citep{Guha2022}, DLAs at $z \sim 1$ \citep[purple histogram with angled lines,][]{Wild2005} and the GOTOQ-subsample (blue histogram with stars) studied here.}
    \label{fig:dust_compare}
\end{figure}

Assuming the \MgII\ absorbing gas follows the dust extinction similar to that of the small magellanic clouds (SMC) \citep{Gordon2003}, we calculate the color excess $\rm{E(B-V)}$ of the background quasar spectra using a standard quasar spectral template \citep{Selsing2016}. We first redshift the quasar template to the redshift of the background quasar and then apply the SMC dust extinction law at the \MgII\ absorption redshift to match it with the shape of the quasar spectrum in hand by varying the dust extinction coefficient, $\rm{A_v}$. The color excess, $\rm E(B - V)$ for our sample ranges from ${-0.191}$ to { 0.422}, with a median value of { 0.058}. The measured values of $\rm E(B-V)$ are provided in the last column of Table \ref{tab:sample}. However, it is important to note that the highest line of sight reddening is observed for the system J1314+0330. As previously mentioned, this system is incorrectly identified as a GOTOQ. Therefore, like before, we do not include it in the analysis presented here.   Given the median $\rm W_{2796}$ of { 2.21 \AA}, the expected color excess towards this GOTOQ sub-sample is ${0.013\pm0.06}$ \citep{Budzynski}. Therefore, the line of sight reddening towards the GOTOQ is significantly higher than the typical \MgII\ absorption systems of similar $\rm W_{2796}$. \citet{Wild2005} have measured the $\rm E(B-V)$ for a sample of Ca~{\sc ii} absorbers. Their values range from $-0.034$ to 0.30 with a median value of 0.042. 

In Figure \ref{fig:dust_compare}, we compare the $\rm E(B-V)$ between the \usmg\ absorption systems at $z \sim 0.5$ from \citet{Guha2022}, Ca~{\sc ii} absorbers from \citet{Wild2005} and the GOTOQ sub-sample studied here. The blue and orange histograms correspond to the $\rm E(B-V)$ distribution of the GOTOQ and the \usmg\ absorption systems, respectively. A KS test between these two distributions yields a $p$-value of { 0.06}, implying that the dust extinctions in these two samples follow similar distributions. This might be expected as one-fourth of the GOTOQs are consistent with being a \usmg\ system. Similarly, the $p$-values obtained while comparing the E(B$-$V) distribution for the GOTOQs and Ca~{\sc ii} systems is { 0.24}. This suggests that E(B$-$V) distributions for these two samples are also statistically similar.

\begin{figure}
    \centering
    \includegraphics[width=0.47\textwidth]{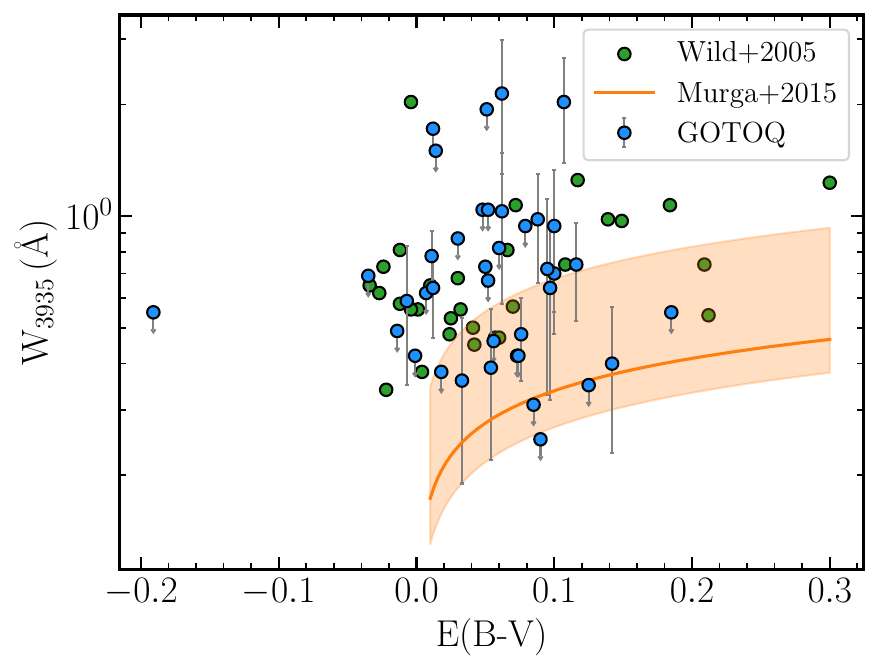}
    \caption{The color excess E(B$-$V) versus the $\rm W^{Ca\textsc{ii}}_{3935}$. Blue points correspond to the GOTOQs studied in this work, and the green points correspond to the Ca~{\sc ii} absorbers from \citet{Wild2005}. The orange line shows the best relation between E(B$-$V) and  $\rm W^{Ca\textsc{ii}}_{3935}$ for our Galaxy and the shaded region shows the corresponding $1\sigma$ error \citep{Murga2015}.}
    \label{fig:wvsebv}
\end{figure}

{ In Figure~\ref{fig:wvsebv}, we plot E(B$-$V) vs. $\rm W_{3935}$ for GOTOQs discussed here, and for the Ca~{\sc ii} absorbers studied by \citet{Wild2005}.  We also compare these measurements of our Galaxy from \citet{Murga2015}. Our measurements for GOTOQs roughly follow the region occupied by the Ca~{\sc ii} absorbers. However, we do not find any significant correlation when considering only our data points.  It is evident from Figure~\ref{fig:wvsebv} that for a given E(B$-$V), the extra-galactic measurements have larger Ca~{\sc ii} rest equivalent width compared to the local measurements by \citet{Murga2015}. While interpreting $\rm W_{3935}$ as an indicator of gas content is debatable, this could imply the dust-to-gas ratio in these absorbers is smaller than seen in our Galaxy. There are several indications for this whenever $\rm E(B-V)$ and $N$(H~{\sc i}) measurements are available at high-z \citep[see for example, discussions in ][]{Ledoux2015, Dutta2020}. Lastly, we investigated the potential correlation between E(B$-$V) and D in our sample and found no statistically significant correlation. The lack of correlation between E(B$-$V) and impact parameter for the low-z GOTOQs was also reported by \citet{Straka_2015}.}

\subsection{Estimation of galaxy parameters}
\begin{figure}
    \centering
    \includegraphics[width=0.45\textwidth]{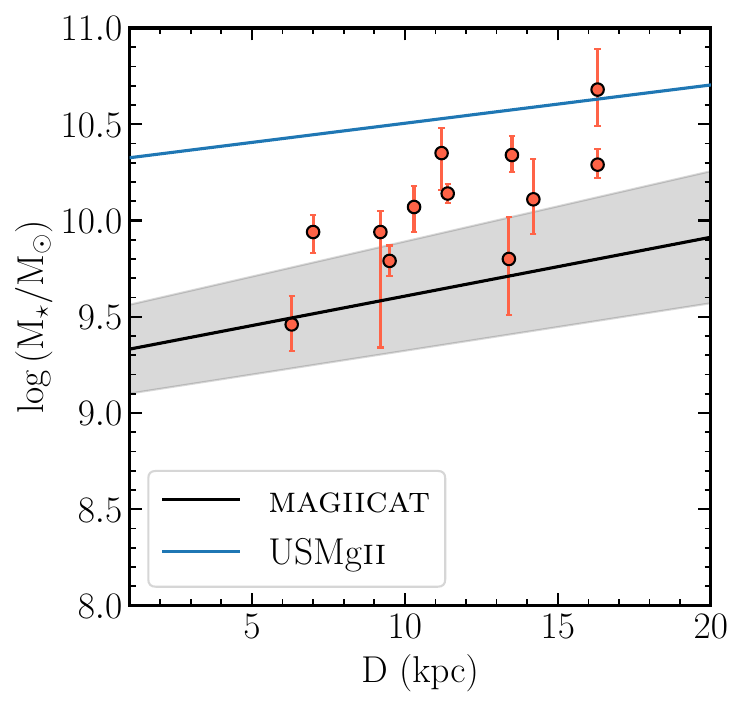}
    \caption{Stellar masses of the foreground galaxies on top of quasars with visible photometric extension versus their impact parameter. The data points are shown in red. The solid black line corresponds to the best-fit line for MAGIICAT galaxies, and the shaded region shows the $1\sigma$ uncertainty to the fit. The blue solid line shows the same for the \usmg\ host galaxies.}
    \label{fig:d_vs_mass}
\end{figure}

\begin{table}
    \centering
    \begin{tabular}{lcr}
        \hline
         Galaxy & $W_{2796}$(\AA) & $\rm{\log\, (M_\star / M_\odot)}$ \\
         (1) & (2) & (3)\\
         \hline
         \hline
         J0129$-$0032 & 3.50$\pm$0.26 &$\rm{10.34^{+0.10}_{-0.09}}$ \\
         J0131$-$0401 & 1.70$\pm$0.12 & $\rm 10.01^{+0.11}_{-0.13}$ \\
         J0235+0059 & 0.93$\pm$0.43 &$\rm{9.46^{+0.15}_{-0.14}}$   \\
         J0807+0945 & $2.12\pm0.21$ & $\rm{10.29^{+0.08}_{-0.07}}$  \\
         J0808+0641 & $1.96\pm0.59$ & $\rm{9.94^{+0.11}_{-0.60}}$ \\
         J0858+0556 & 2.90$\pm$1.03 & $\rm{9.90}^{+0.09}_{-0.11}$ \\
         J0929$-$0126 & $1.47\pm0.43$ & $\rm{9.80^{+0.22}_{-0.29}}$  \\
         {J1218+0309} & 2.80$\pm$0.39 & $\rm{10.14}^{+0.05}_{-0.05}$ \\
         J1354+0756 & 2.15$\pm$0.31 &$\rm{10.11^{+0.21}_{-0.18}}$ \\
         J1356+0601 & 2.52$\pm$0.19 &$\rm{10.35^{+0.13}_{-0.19}}$ \\
         J2244$-$0215 & 3.77$\pm$0.20 &$\rm{10.68^{+0.21}_{-0.19}}$ \\
         J2342+0101 & 2.35$\pm$0.27 & $\rm 9.79^{+0.08}_{-0.08}$ \\
        \hline

    \end{tabular}
    \caption{Measurement of Stellar Mass in the case of GOTOQs where accurate photometric measurements are possible.  Column (1): Galaxy name, Column (2): $W_{2796}$, and Column (3): the measured value of Stellar mass.}
    \label{tab:gal_props}
\end{table}

There is a correlation found between stellar mass $\rm M_\star$ and impact parameter for the general population of Mg~{\sc ii} absorbers in \citet[]{Guha2022}. It has also been found that for a given impact parameter, the  $\rm M_\star$ estimated for the \usmg\ absorbers tend to be higher compared to the Mg~{\sc ii} absorbers with low $\rm W_{2796}$. These correlations imply that (i) at a given impact parameter, the absorbers with large $\rm W_{2796}$ tend to have large $\rm M_\star$ and (ii) for a given $\rm W_{2796}$ galaxies at higher impact parameters tend to have higher values of $\rm M_\star$. Therefore, if GOTOQs follow this trend, we expect the $\rm M_\star$ for the host galaxies to be typically less than what has been found for \usmg\ host galaxies at high impact parameters. Indeed, \citet{Guha2023} have measured  $\rm 9.4 \leqslant log (M_\star/M_\odot) \leqslant 9.8$ over the redshift range $\rm 0.35\leqslant z\leqslant 1.1$. Here, we check whether such measurements for individual cases in the case of GOTOQs follow the above-mentioned trend.

Accurate photometric measurements are required in several bands to measure the stellar masses ($M_\star$) of GOTOQs using SED fitting techniques. This is possible only for { 12} of the {39} objects with clear photometric extensions in our sample. For these twelve systems, the angular separations between the quasars and the foregrounds are large, so reliable fluxes in different photometric bands can be measured by subtracting the quasar PSF. We computed the foreground galaxy fluxes in the \emph{grz} bands of the DECaLS photometry. Then, we fitted the spectral energy distributions (SED) using the publicly available code \texttt{BAGPIPES} \citep{Carnall2018}. During the fit, we assumed a delayed star formation history \citep{Carnall2019} and kept the redshift constant as obtained from the \OII\ emission line. Obtained stellar masses of the GOTOQ with photometric counterparts are given in column 3 of Table \ref{tab:gal_props}.  In Figure~\ref{fig:d_vs_mass}, we show plot $\rm M_\star$ vs. D for these objects and also show the best fit obtained for the host galaxies of \usmg\ and that obtained for the MAGIICAT galaxies. { However, it is important to note that the correlation observed for the GOTOQs may be influenced by the mass-size relation of galaxies \citep{Ward2024}. Since more massive galaxies tend to be larger, massive GOTOQs are also larger in size and, consequently, may have larger impact parameters.} It will be good to measure the host galaxy properties for more GOTOQs using HST photometry or AO-assisted ground-based imaging.  This will provide more insights into how stellar mass controls the equivalent width distributions at low impact parameters.

\section{Summary}
\label{sec:discussions}

In this work, we present long-slit spectroscopic observations of { 40} GOTOQs having \OII\ luminosity above ${4.2\times10^{40} ergs s^{-1}}$ over the redshift range ${0.37\leqslant z \leqslant 1.01}$.  The \MgII\ rest equivalent width of our sample spans ${0.93 \text{\AA} \leqslant W_{2796}\leqslant 4.75}$\AA\ with a median value of { 2.21\AA}.

We are successful in measuring the impact parameters for { 31} GOTOQs.  The impact parameter ranges from { 3.6 to 16.3 kpc} with a median impact parameter of { 8.6 kpc}. In { 19} cases, we have simultaneous impact parameter measurements using our spectroscopic observations with SALT and photometric observations from DECaLS. Apart from the two cases, they agree very well between them within an uncertainty of $\sim$ 1.5 kpc. This reassures that the residual fluxes measured around quasars with correct photometric redshifts are the host galaxies of GOTOQs. The measured impact parameters also confirm that the \OII\ luminosity measured using SDSS fiber spectra is severely affected by fiber losses, particularly the BOSS spectra.

We do not find any correlation or anti-correlation between $\rm W_{2796}$ and D.  This aligns with our earlier finding that these two quantities are uncorrelated at small impact parameters. Based on all the available spectroscopically confirmed host galaxy measurements, we provide the best-fitted log-linear relationship between these two quantities. 

We report the detection of Ca~{\sc ii} $\lambda$3935 and Mn~{\sc ii} $\lambda$2576 towards {16} and 17 GOTOQs, respectively.  For $W_{3935} \geqslant 0.7$\AA\ (i.e., strong Ca~{\sc ii} absorbers by usual definition), we find the covering factor to be 0.23 around GOTOQs. We do not find any correlation between $\rm W_{3935}$ and D. This is consistent with the finding of \citet{Straka_2015} for low-z GOTOQs. It is usually conjectured that weak Ca~{\sc ii} absorbers originate mainly from the halo gas, while strong Ca~{\sc ii} absorbers originate from the disk and/or inner halo gas. In such a case, we expect an anti-correlation between $W_{3935}$ and D. The Absence of this correlation could be related to the lack of measurements from absorbers at higher D having low $W_{3935}$. We also explore the correlation between REW of Mn~{\sc ii} $\lambda$2576, Fe~{\sc ii} $\lambda$2600 and Mg~{\sc i} $\lambda$2852. We do not find any statistically significant correlation or anti-correlation with D for REW of these ions.

The line of sight reddening (quantified using E(B$-$V)) obtained using the SMC extinction curve for GOTOQs is statistically similar for high-z \MgII\ systems.  They are also found to be similar to what has been measured for Ca~{\sc ii} selected absorbers studied by \citet{Wild2005}. In the E(B$-$V) vs $W_{3935}$ plane, GOTOQs occupy the same region as Ca~{\sc ii} absorbers of \citet{Wild2005}. However, for a given E(B$-$V), both the sample show larger $W_{3935}$ than what has been found in our Galaxy by \citet{Murga2015}. This can be attributed to the smaller dust-to-gas ratio in high-$z$ absorbers. We do not find any correlation between E(B$-$V) and impact parameters. This is also consistent with the result of \citet{Straka_2015} for low-z GOTOQs.

Measurement of $M_\star$ is possible for { 12} GOTOQs in our sample. As in the case of the general population of Mg~{\sc ii} absorbers, these measurements are consistent with (i) at a given impact parameter, the absorbers with large $W_{2796}$ tend to have large $M_\star$ and (ii) for a given $W_{2796}$ galaxies at higher impact parameters tend to have higher values of $M_\star$. Establishing the relationship between these quantities using a large sample is essential for understanding how stellar mass controls the equivalent width distributions at low impact parameters.

\section*{Data Availability}
Data used in this work are obtained using SALT. Raw data will become available for public use 1.5 years after the observing date at https://ssda.saao.ac.za/.

\section*{Acknowledgement}
\label{sec:acknowledge}

This project makes use of the following softwares : NumPy \citep{numpy2020}, SciPy \citep{scipy2020}, Matplotlib \citep{matplotlib2007}, AstroPy \citep{astropy:2013, astropy:2018}, and Ultranest \citep{Buchner2021}.

All the new observations reported in this paper were obtained with the Southern African Large Telescope (SALT).

This paper makes use of SDSS observational data. Funding for the Sloan Digital Sky Survey IV has been provided by the Alfred P. Sloan Foundation, the U.S. Department of Energy Office of Science, and the Participating Institutions. SDSS-IV acknowledges support and resources from the Center for High-Performance Computing at the University of Utah. The SDSS website is www.sdss.org. SDSS-IV is managed by the Astrophysical Research Consortium for the Participating Institutions of the SDSS Collaboration, including  the Brazilian Participation Group, the Carnegie Institution for Science, Carnegie Mellon University, Center for Astrophysics | Harvard \& Smithsonian, the Chilean Participation 
Group, the French Participation Group, Instituto de Astrof\'isica de Canarias, The Johns Hopkins University, Kavli Institute for the  Physics and Mathematics of the Universe (IPMU) / University of Tokyo, the Korean Participation Group, Lawrence Berkeley National Laboratory, Leibniz Institut f\"ur Astrophysik Potsdam (AIP),  Max-Planck-Institut  f\"ur Astronomie (MPIA Heidelberg), Max-Planck-Institut f\"ur Astrophysik (MPA Garching), Max-Planck-Institut f\"ur Extraterrestrische Physik (MPE), National Astronomical Observatories of China, New Mexico State University, New York University, University of Notre Dame, Observat\'ario Nacional / MCTI, The Ohio State University, Pennsylvania State University, Shanghai Astronomical Observatory, United Kingdom Participation Group, 
niversidad Nacional Aut\'onoma de M\'exico, University of Arizona, University of Colorado Boulder, University of Oxford, University of Portsmouth, University of Utah, University of Virginia, University of Washington, University of Wisconsin, Vanderbilt University, and Yale University.

The DESI Legacy Imaging Surveys consist of three individual and complementary projects: the Dark Energy Camera Legacy Survey (DECaLS), the Beijing-Arizona Sky Survey (BASS), and the Mayall z-band Legacy Survey (MzLS). DECaLS, BASS, and MzLS together include data obtained, respectively, at the Blanco telescope, Cerro Tololo Inter-American Observatory, NSF’s NOIRLab; the Bok telescope, Steward Observatory, University of Arizona; and the Mayall telescope, Kitt Peak National Observatory, NOIRLab. NOIRLab is operated by the Association of Universities for Research in Astronomy (AURA) under a cooperative agreement with the National Science Foundation. Pipeline processing and analyses of the data were supported by NOIRLab and the Lawrence Berkeley National Laboratory (LBNL). Legacy Surveys also uses data products from the Near-Earth Object Wide-field Infrared Survey Explorer (NEOWISE), a project of the Jet Propulsion Laboratory/California Institute of Technology, funded by the National Aeronautics and Space Administration. Legacy Surveys was supported by: the Director, Office of Science, Office of High Energy Physics of the U.S. Department of Energy; the National Energy Research Scientific Computing Center, a DOE Office of Science User Facility; the U.S. National Science Foundation, Division of Astronomical Sciences; the National Astronomical Observatories of China, the Chinese Academy of Sciences and the Chinese National Natural Science Foundation. LBNL is managed by the Regents of the University of California under contract to the U.S. Department of Energy. The complete acknowledgments can be found at https://www.legacysurvey.org/acknowledgment/. The Photometric Redshifts for the Legacy Surveys (PRLS) catalog used in this paper was produced thanks to funding from the U.S. Department of Energy Office of Science, Office of High Energy Physics via grant DE-SC0007914.



\bibliographystyle{mnras}
\bibliography{example} 

\bsp	
\label{lastpage}
\end{document}